\documentclass[prd,aps,superscriptaddress,amsmath,amssymb,nofootinbib,longbibliography,twocolumn,eqsecnum]{revtex4-1}

\usepackage{graphicx}       

\usepackage{color}
\usepackage{drafttl}

\newcommand{\realn}{{\mathbb{R}}}

\newcommand{\ts}[1]{{\boldsymbol{#1}}}         

\newcommand{\calA}{\mathcal{A}}
\newcommand{\calJ}{\mathcal{J}}
\newcommand{\calU}{\mathcal{U}}

\newcommand{\barA}{\bar{A}}
\newcommand{\barJ}{\bar{J}}

\newcommand{\barcalA}{\bar{\mathcal{A}}}
\newcommand{\barcalJ}{\bar{\mathcal{J}}}
\newcommand{\barcalU}{\bar{\mathcal{U}}}

\newcommand{\tilA}{\tilde{A}}
\newcommand{\tilB}{\tilde{B}}
\newcommand{\tilJ}{\tilde{J}}
\newcommand{\tilU}{\tilde{U}}
\newcommand{\tilX}{\tilde{X}}
\newcommand{\tilDel}{\tilde{\Delta}}
\newcommand{\tilSig}{\tilde{\Sigma}}

\newcommand{\tilcalA}{\tilde{\mathcal{A}}}
\newcommand{\tilcalJ}{\tilde{\mathcal{J}}}
\newcommand{\tilcalU}{\tilde{\mathcal{U}}}

\newcommand{\bartilA}{\tilde{\bar{A}}}
\newcommand{\bartilJ}{\tilde{\bar{J}}}

\newcommand{\btilcalA}{\tilde{\bar{\mathcal{A}}}}
\newcommand{\btilcalJ}{\tilde{\bar{\mathcal{J}}}}
\newcommand{\btilcalU}{\tilde{\bar{\mathcal{U}}}}

\newcommand{\tilx}{\tilde{x}}
\newcommand{\tilphi}{\tilde{\phi}}

\newcommand{\framevec}[1]{\frac{\bpart}{\bpart #1}} 

\newcommand{\bd}{\mathbf{d}}
\newcommand{\be}{\boldsymbol{e}}
\newcommand{\bg}{\boldsymbol{g}}
\newcommand{\bh}{\boldsymbol{h}}
\newcommand{\bk}{\boldsymbol{k}}
\newcommand{\bl}{\boldsymbol{l}}
\newcommand{\br}{\boldsymbol{r}}
\newcommand{\bs}{\boldsymbol{s}}
\newcommand{\bpart}{\boldsymbol{\partial}}
\newcommand{\beps}{\boldsymbol{\epsilon}}
\newcommand{\bpi}{\boldsymbol{\pi}}
\newcommand{\bxi}{\boldsymbol{\xi}}
\newcommand{\bomega}{\boldsymbol{\omega}}

\newcommand{\hatbeps}{\hat{\boldsymbol{\epsilon}}}

\newcommand{\tilPhi}{\tilde{\boldsymbol{\Phi}}}

\newcommand{\tilk}{\tilde{\boldsymbol{k}}}
\newcommand{\tilK}{\tilde{\boldsymbol{K}}}
\newcommand{\till}{\tilde{\boldsymbol{l}}}
\newcommand{\tilL}{\tilde{\boldsymbol{L}}}
\newcommand{\tilr}{\tilde{\boldsymbol{r}}}
\newcommand{\tilR}{\tilde{\boldsymbol{R}}}
\newcommand{\tils}{\tilde{\boldsymbol{s}}}

\newcommand{\tilbg}{\tilde{\boldsymbol{g}}}
\newcommand{\tilbeps}{\tilde{\boldsymbol{\epsilon}}}
\newcommand{\tilpi}{\tilde{\boldsymbol{\pi}}}
\newcommand{\tilxi}{\tilde{\boldsymbol{\xi}}}
\newcommand{\tilXi}{\tilde{\boldsymbol{\Xi}}}

\newcommand{\htilbeps}{\hat{\tilde{\boldsymbol{\epsilon}}}}

\newcommand{\barN}{{\bar{N}}}
\newcommand{\baral}{{\bar{\alpha}}}
\newcommand{\barbe}{{\bar{\beta}}}
\newcommand{\barmu}{{\bar{\mu}}}
\newcommand{\barnu}{{\bar{\nu}}}

\newcommand{\tilN}{\tilde{N}}
\newcommand{\tila}{\tilde{a}}
\newcommand{\tilb}{\tilde{b}}
\newcommand{\tilmu}{\tilde{\mu}}

\newcommand{\bartilN}{\tilde{\bar{N}}}

\newcommand{\upind}[1]{{}^{#1}\!} 
\newcommand{\upal}{\upind{\alpha}}
\newcommand{\upbe}{\upind{\beta}}
\newcommand{\upga}{\upind{\gamma}}

\begin{document}

\title{Higher-dimensional black holes with multiple equal rotations}

\author{Eli\v{s}ka Pol\'a\v{s}kov\'a}
\email{eli.polaskova@email.cz}
\author{Pavel Krtou\v{s}}
\email{Pavel.Krtous@utf.mff.cuni.cz}
\affiliation{Institute of Theoretical Physics,
Faculty of Mathematics and Physics, Charles University,
V~Hole\v{s}ovi\v{c}k\'ach~2, Prague, 18000, Czech Republic}

\date{November 4, 2021}  

\begin{abstract}
We study a limit of the Kerr--(A)dS spacetime in a general dimension where an arbitrary number of its rotational parameters is set equal. The resulting metric after the limit formally splits into two parts --- the first part has the form of the Kerr--NUT--(A)dS metric analogous to the metric of the entire spacetime, but only for the directions not subject to the limit, and the second part can be interpreted as the K\"{a}hler metrics. However, this separation is not integrable, thus it does not lead to a product of independent manifolds. We also reconstruct the original number of explicit and hidden symmetries associated with Killing vectors and Killing tensors. Therefore, the resulting spacetime represents a special subcase of the generalized Kerr--NUT--(A)dS metric that retains the full Killing tower of symmetries. In $D=6$, we present evidence of an enhanced symmetry structure after the limit. Namely, we find additional Killing vectors and show that one of the Killing tensors becomes reducible as it can be decomposed into Killing vectors.
\end{abstract}

\maketitle


\section{Introduction}
\label{sc:intro}

Four-dimensional black holes have been widely studied for more than a hundred years. Nowadays, they are used as astrophysically relevant sources which led to several breakthrough observations in recent years. These include the first detection of gravitational waves from a binary black hole merger \cite{Abbott2016}, the first image of the supermassive black hole candidate in the center of the galaxy M87, made by the Event Horizon Telescope \cite{EHT1,EHT6}, or the explanation of the star motion near the black hole in the centre of our galaxy \cite{Genzel2010,Ghez2008}. The standard model of the black hole used in such astrophysical situations is the Kerr solution of the Einstein equations in four-dimensional general relativity.

From the mathematical point of view, Kerr black holes \cite{Kerr1963} are included in a large family of solutions known as the Pleba\'{n}ski--Demia\'{n}ski metric \cite{Plebanski1976}. This metric represents spacetimes of algebraic type D that solve the vacuum Einstein equations with the cosmological constant, and it is characterized by seven arbitrary parameters, which can be interpreted as the cosmological constant, mass, NUT parameter, rotation, acceleration, electric and magnetic charge. It contains many well-known spacetimes as special cases --- apart from the Kerr metric, which describes an axially symmetric rotating black hole, it also includes for example the Taub--NUT (Newman--Unti--Tamburino) solution \cite{Taub1951,Newman1963} with one NUT parameter as well as accelerating black holes represented by the C-metric \cite{Griffiths2009}.

Unlike the Kerr spacetime, the Taub--NUT metric does not have such a clear physical interpretation --- the presence of a NUT parameter in a four-dimensional spacetime leads to pathologies such as the existence of closed timelike curves \cite{Misner1963}. However, some of these pathologies can be regarded as an unphysical feature of the idealized inner solution, which disappears when a realistic matter source for the outer solution is introduced.

In this work, we study generalization of these black holes to higher dimensions. The motivation for studying higher-dimensional metrics in general is their connection with string theory, the AdS/CFT correspondence and brane-world models. Moreover, the perspective of a general dimension may deepen the understanding of studied solutions. Last but not least, they are interesting from the mathematical point of view. An extensive review of higher-dimensional black hole solutions can be found in \cite{Emparan2008}.

One of the interesting higher-dimensional solutions, which generalizes the black hole solutions known in four dimensions, is called the \emph{Kerr--NUT--(A)dS metric} \cite{Chen2006}. It is characterized by the cosmological constant, mass, rotational and NUT parameters, however, it does not include acceleration and electric/magnetic charge. Therefore, a generalization of the Pleba\'{n}ski--Demia\'{n}ski metric to higher dimensions is yet to be discovered. The Kerr--NUT--(A)dS metric can describe various geometries of both the Euclidean and the Lorentzian signature, such as maximally symmetric spaces, so-called Euclidean instantons, and black holes. It also includes well-known higher-dimensional solutions as special cases, for example the Myers--Perry black hole \cite{Myers1986} (generalization of the Kerr black hole), the Kerr--(A)dS metric \cite{Gibbons2004, Gibbons2005} (generally rotating black hole in an asymptotically (anti)-de Sitter spacetime) and the higher-dimensional Taub--NUT--(A)dS metric \cite{Mann2004,Mann2006}.

Higher-dimensional rotating black holes display many similar properties to their four-dimensional counterparts. This is caused by the fact that these spacetimes admit a special geometrical object, which we refer to as the \emph{principal tensor} \cite{Frolov2007,Houri2007,Kubiznak2007, Krtous2008}. It is defined as a non-degenerate closed conformal Killing--Yano tensor.

The very existence of the principal tensor significantly restricts the geometry --- the most general geometry consistent with the existence of this tensor is the off-shell Kerr--NUT--(A)dS geometry. Here, the attribute ``off-shell'' refers to a general form of the metric that does not require the vacuum Einstein equations. The principal tensor generates a rich symmetry structure called the Killing tower \cite{Krtous2007,Frolov2008a}, which includes Killing vectors and Killing tensors associated with explicit and hidden symmetries of the spacetime.

Moreover, it uniquely determines canonical coordinates in which the Hamilton--Jacobi \cite{Frolov2003} and the Klein--Gordon equations \cite{Frolov2007a,Sergyeyev2008,Kolar2016} as well as the Dirac \cite{Oota2008,Cariglia2011,Cariglia2011a} and the Maxwell equations \cite{Lunin:2017,Frolov:2018ezx,Krtous:2018,Frolov:2018a} are fully separable, and the geodesic motion is completely integrable \cite{Krtous2007,Page2007,Krtous2007a}. Separability has been demonstrated also for higher-form fields \cite{Lunin2019}. As one can see, the principal tensor indeed plays a very important role in higher-dimensional black hole physics. For an extensive review of the role of the principal tensor and other properties of the Kerr--NUT--(A)dS geometry, see \cite{Frolov2017a}.

Apart from the Kerr--NUT--(A)dS spacetime and its properties, several limit cases of the general metric were also studied, such as the near-horizon limits \cite{Lu2009,Chernyavsky2014,Xu2015,Galajinsky2016}. Furthermore, the limit where some of the black hole's rotations are switched off was investigated \cite{Krtous2016a}. Such a limit leads to warped spaces deformed and twisted by the NUT parameters, which thus do not maintain their unphysical properties when present in a space with the Euclidean signature. Another limit case where particular roots of the metric functions degenerate was studied \cite{Kolar2017}, which results in geometries such as the Taub--NUT--(A)dS metric and the extreme near-horizon geometry.

These papers have thus demonstrated that not only can performing various limits of the general metric shed light on the role of various metric parameters, but it can also lead to new interesting geometries. Moreover, the resulting spacetimes are expected to possess an enhanced symmetry structure after the limit, which is manifested in the presence of additional Killing vectors and also in the reducibility of Killing tensors that can be decomposed into Killing vectors. Reducibility properties of Killing tensors were also studied in four dimensions for near-horizon geometries \cite{Galajinsky2010,Galajinsky2011}.

However, performing a limiting procedure is not in general a trivial task since certain regions of the spacetime can shrink or expand during the limit and become degenerate. Therefore, it is usually necessary to accompany the limiting procedure by a suitable rescaling of coordinates and parameters.

This work is focused on a particular limit case of the general Kerr--NUT--(A)dS metric, namely, the \emph{equal-spin limit}. It is the limit where an arbitrary number of rotational para\-meters of the spacetime coincides.

The paper is organized as follows. In Section \ref{sc:kerr-nut-ads}, which is an overview of already known results, we introduce the Kerr--NUT--(A)dS spacetime and summarize its properties. The next two sections are dedicated to a general equal-spin limit. Namely, in Section \ref{sc:metric-limit} we introduce the parametrization of the limit and apply it to the metric, while Section \ref{sc:symmetries-limit} discusses the limit form of the principal tensor, Killing vectors and Killing tensors. Section \ref{sc:examples} presents explicit examples of the general results obtained in Section \ref{sc:metric-limit} and \ref{sc:symmetries-limit} --- it focuses on black holes with all the rotational parameters set equal. Additional technical results and detailed calculations are provided in the appendices. Appendix \ref{app:metric-funcs-def} summarizes definitions and useful identities concerning auxiliary functions that appear in the metric before the limit. Appendix \ref{app:metric-funcs-limit} provides the limit form of these auxiliary functions.


\section{Kerr--NUT--(A)dS geometry}
\label{sc:kerr-nut-ads}

\subsection{Canonical form of the metric}

The off-shell Kerr--NUT--(A)dS metric in ${D=2N}$ dimensions\footnote{For simplicity, we restrict ourselves to even dimensions. The generalization to odd dimensions is possible --- a corresponding term must be added to the metric and other related quantities. Otherwise, the analysis remains the same for both cases.}, which is consistent with the existence of the principal tensor, can be written in the form \cite{Frolov2017a}
\begin{equation}
\label{eq:metric-phi}
	\bg = \sum_{\mu}\left[\frac{U_{\mu}}{X_{\mu}}\bd x_{\mu}^2 + \frac{X_{\mu}}{U_{\mu}}\left(\sum_{\nu}\frac{J_{\mu}(a_{\nu}^2)}{\lambda a_{\nu}\calU_{\nu}} \bd \phi_{\nu}\right)^2\right] \,,
\end{equation}
with Greek indices\footnote{The Einstein summation convention is not used for these indices. Also, we do not indicate their ranges explicitly in sums or products, unless they differ from the default above.} going over the range
\begin{equation}
	\mu, \nu, \ldots = 1, \ldots, N \,.
\end{equation}

The metric functions ${X_\mu}$ should be arbitrary functions of a single coordinate ${x_\mu}$,  $X_{\mu}=X_{\mu}(x_{\mu})$. The metric functions $U_{\mu}$ and $J_{\mu}(a_{\nu}^2)$ are polynomials in all coordinates $x_{\nu}$
\begin{equation}
\label{eq:funcUJmu}
	U_{\mu} = \prod_{\overset{\scriptstyle{\nu}}{\scriptstyle{\nu\neq\mu}}}\left(x_{\nu}^2-x_{\mu}^2\right)\,, \qquad
	J_{\mu}(a_{\nu}^2) = \prod_{\overset{\scriptstyle{\kappa}}{\scriptstyle{\kappa\neq\mu}}}\left(x_{\kappa}^2-a_{\nu}^2\right) \,,
\end{equation}
and the functions $\calU_{\mu}$ of the metric parameters $a_\mu$ are defined similarly to $U_{\mu}$, only with $x_{\mu}$ replaced by $a_{\mu}$.

When the ${\Lambda}$-vacuum Einstein equations are imposed, we obtain the on-shell Kerr--NUT--(A)dS geometry. The metric functions ${X_\mu}$ must be polynomials in the form
\begin{equation}
\label{eq:funcX}
	X_{\mu} = \lambda\calJ(x_{\mu}^2)-2b_{\mu}x_{\mu} \,,
\end{equation}
where $\calJ(x_{\mu}^2)$ reads
\begin{equation}
\label{eq:funcJ}
	\calJ(x_{\mu}^2) = \prod_{\nu}\left(a_{\nu}^2-x_{\mu}^2\right) \,.
\end{equation}
The complete list of metric functions as well as important relations between them can be found in Appendix \ref{app:metric-funcs-def}.

It is useful to introduce the following orthogonal frames of 1-forms
\begin{equation}
\label{eq:OG-frame}
\begin{split}
	\be^{\mu} = \left(\frac{U_{\mu}}{X_{\mu}}\right)^{\frac{1}{2}}\beps^{\mu} & = \left(\frac{U_{\mu}}{X_{\mu}}\right)^{\frac{1}{2}}\bd x_{\mu} \,, \\[0.1cm]
	\hat{\be}^{\mu} = \left(\frac{X_{\mu}}{U_{\mu}}\right)^{\frac{1}{2}}\hat{\beps}^{\mu} & = \left(\frac{X_{\mu}}{U_{\mu}}\right)^{\frac{1}{2}} \sum_{\nu}\frac{J_{\mu}(a_{\nu}^2)}{\lambda a_{\nu}\calU_{\nu}}\bd\phi_{\nu} \,,
\end{split}
\end{equation}
where \{$\be^{\mu}$, $\hat{\be}^{\mu}$\} is normalized and \{$\beps^{\mu}$, $\hat{\beps}^{\mu}$\} is not normalized. Similarly, dual orthogonal frames of vectors read
\begin{equation}
\label{eq:OG-coframe}
\begin{split}
	\be_{\mu} = \left(\frac{X_{\mu}}{U_{\mu}}\right)^{\frac{1}{2}}\beps_{\mu} & = \left(\frac{X_{\mu}}{U_{\mu}}\right)^{\frac{1}{2}}\framevec{x_{\mu}} \,, \\[0.1cm]
	\hat{\be}_{\mu} = \left(\frac{U_{\mu}}{X_{\mu}}\right)^{\frac{1}{2}}\hat{\beps}_{\mu} & = \left(\frac{U_{\mu}}{X_{\mu}}\right)^{\frac{1}{2}} \sum_{\nu}\frac{\lambda a_{\nu}\calJ_{\nu}(x_{\mu}^2)}{U_{\mu}} \framevec{\phi_{\nu}} \,.
\end{split}
\end{equation}
Using these frames, the metric can be written simply as
\begin{equation}
\label{eq:metric-frame}
	\bg = \sum_{\mu}\left(\be^{\mu}\be^{\mu} + \hat{\be}^{\mu}\hat{\be}^{\mu}\right) = \sum_{\mu} \left(\frac{U_{\mu}}{X_{\mu}}\beps^{\mu}\beps^{\mu} + \frac{X_{\mu}}{U_{\mu}}\hatbeps^{\mu}\hatbeps^{\mu}\right) \,.
\end{equation}

The coordinates we used are divided into two sets, $x_{\mu}$ and $\phi_{\mu}$. Since the metric functions are independent of $\phi_{\mu}$, they are the Killing coordinates. Their corresponding Killing vectors have fixed points and thus define the axes of rotational symmetry \cite{Kolar2019}. In the black hole case (i.e. for the Lorentzian signature), they are also related to the temporal coordinate. The coordinates $x_{\mu}$ represent radius and latitudinal angles.

Alternatively, the metric can be expressed using another set of angular coordinates $\psi_k$, $k=0,\ldots,N-1$, instead of $\phi_{\mu}$. The coordinates $\psi_k$ are suitable for constructing the Killing tower and studying explicit and hidden symmetries. However, $\phi_{\mu}$ are better suited for the physical interpretation of the metric. An explicit definition of $\psi_k$ coordinates and the corresponding form of the metric can be found, e.g., in \cite{Frolov2017a}.

The on-shell metric is described by the parameters $a_{\mu}$, $b_{\mu}$ and $\lambda$, where $\lambda$ is related to the cosmological constant $\Lambda$ as
\begin{equation}
	\Lambda = (2N-1)(N-1)\lambda \,.
\end{equation}
In general, the parameters $a_{\mu}$ are related to rotations and the parameters $b_{\mu}$ encode mass and NUT charges. However, when the NUT charges are non-vanishing, the interpretation of parameters is not so straightforward \cite{Frolov2017a,Kolar2019}.

\subsection{Black hole}

The general metric \eqref{eq:metric-phi} can have both the Euclidean and the Lorentzi\-an signature, depending on our choice of the coordinate ranges and values of the parameters. Detailed discussion of geometries with the Euclidean signature, such as maximally symmetric spaces and Euclidean instantons, can be found in \cite{Frolov2017a}.

The Lorentzian signature can be obtained by Wick-rotating the following coordinates and parameters
\begin{equation}
\label{eq:Wick}
	x_N = ir \,, \qquad \phi_N = \lambda a_N t \,, \qquad b_N = iM \,,
\end{equation}
where $t$, $r$ and $M$ acquire real values. Moreover, we use a one-parametric gauge freedom in rescaling the parameters and set
\begin{equation}
\label{eq:gauge}
	a_N^2 = -\frac{1}{\lambda} \,.
\end{equation}
We also assume that $a_{\barmu}$ are ordered as
\begin{equation}
	0 < a_1 < \ldots < a_{\barN-1} < a_{\barN} \,,
\end{equation}
where barred indices go over the range
\begin{align}
	\barmu, \barnu, \ldots & = 1, \ldots, \barN \,,\\[0.1cm]
    \barN & = N-1 \,.
\end{align}
The coordinate ${x_\barmu}$ acquires values between the roots of the metric functions ${X_\barmu}$, see \cite{Frolov2017a} for details.

In case of vanishing NUT charges and non-zero mass, i.e. $b_{\barmu}=0$, the coordinate ranges reduce to
\begin{equation}
    a_{\barmu-1}<x_\barmu<a_\barmu\;,
\end{equation}
with the exception of ${-a_1<x_1<a_1}$. In this case, we can introduce $\barN+1$ coordinates $\mu_0$, $\mu_{\barnu}$ instead of $\barN$ coordinates $x_{\bar{\kappa}}$ using the Jacobi transformation in the form\footnote{Barred metric functions are defined in the same way as their unbarred counterparts, only with modified sets of coordinates and parameters (i.e. without $x_N$ and $a_N$).}
\begin{equation}
\label{eq:jacobi-mu}
	\mu_{\barnu}^2 = \frac{\barJ(a_{\barnu}^2)}{-a_{\barnu}^2\barcalU_{\barnu}} \,,\qquad
	\mu_0^2 = \frac{\barJ(0)}{\barcalJ(0)} = \frac{\barA^{(\barN)}}{\barcalA^{(\barN)}} \,,
\end{equation}
which satisfy the constraint
\begin{equation}
\label{eq:sphere-mu}
	\sum_{\barnu=0}^{\barN}\mu_{\barnu}^2 = 1 \,.
\end{equation}
The set \{$t$, $r$, $\mu_0$, $\mu_{\barnu}$, $\phi_{\barnu}$\} is also known as the Myers--Perry coordinates. The Kerr--(A)dS metric in these coordinates is then \cite{Gibbons2004, Gibbons2005}
\begin{equation}
\label{eq:metric-mu}
\begin{split}
	\bg = & -\left(1-\lambda R^2\right) \bd t^2 \\[0.25cm]
	& + \frac{2Mr}{\Sigma} \left[\bd t + \sum_{\barnu} \frac{a_{\barnu}\mu_{\barnu}^2}{1+\lambda a_{\barnu}^2} \left(\bd\phi_{\barnu}-\lambda a_{\barnu}\bd t\right)\right]^2 \\[0.25cm]
	& + \frac{\Sigma}{\Delta_r}\bd r^2 + r^2\bd\mu_0^2 + \sum_{\barnu} \frac{r^2+a_{\barnu}^2}{1+\lambda a_{\barnu}^2} \left( \bd\mu_{\barnu}^2 + \mu_{\barnu}^2 \bd\phi_{\barnu}^2 \right) \\[0.1cm]
	& + \frac{\lambda}{1-\lambda R^2} \left( r^2\mu_0\bd\mu_0 + \sum_{\barnu} \frac{r^2+a_{\barnu}^2}{1+\lambda a_{\barnu}^2} \mu_{\barnu}\bd\mu_{\barnu} \right)^2 \,,
\end{split}
\end{equation}
where
\begin{equation}
\label{eq:bh-func-mu}
\begin{split}
	1-\lambda R^2 & = \left(1-\lambda r^2\right) \left(\mu_0^2 + \sum_{\barnu} \frac{\mu_{\barnu}^2}{1+\lambda a_{\barnu}^2}\right) \,,\\[0.35cm]
	\Delta_r & = \left(1-\lambda r^2\right) \prod_{\barnu} \left(r^2+a_{\barnu}^2\right) - 2Mr \,,\\[0.2cm]
	\Sigma & = \left(\mu_0^2 + \sum_{\barnu} \frac{r^2\mu_{\barnu}^2}{r^2 + a_{\barnu}^2}\right) \prod_{\barmu}\left(r^2 + a_{\barmu}^2\right) \,.
\end{split}
\end{equation}

Finally, for zero NUT charges $b_{\barmu}=0$, the parameters $a_{\barmu}$ can be directly identified with the rotational parameters of the black hole.

\subsection{Explicit and hidden symmetries}

The Kerr--NUT--(A)dS spacetime possesses symmetries of two kinds --- explicit symmetries, which are represented by Killing vector fields, and hidden symmetries, which in our case will be described by Killing tensors. There exists an object behind the symmetry structure --- the principal tensor. It is a crucial object that does not only uniquely determine the canonical form of the metric \cite{Chen2006}, but it also generates the entire tower of Killing vectors and Killing tensors.

The principal tensor is defined as a closed conformal Killing--Yano 2-form that is also non-degenerate (it has functionally independent non-constant eigenvalues). In terms of the coordinates introduced above, it can be written as
\begin{equation}
\label{eq:h-explicit}
	\bh = \sum_{\mu} x_{\mu}\be^{\mu}\wedge\hat{\be}^{\mu}
\end{equation}
using the orthonormal frame \{$\be^{\mu}$, $\hat{\be}^{\mu}$\} \eqref{eq:OG-frame}.

The Killing tower of explicit and hidden symmetries can be constructed from the principal tensor either directly \cite{Krtous2007, Frolov2008, Frolov2008a} or using generating functions \cite{Frolov2017a}. In this work, we will use the method of generating functions.

This method introduces auxiliary $\beta$-dependent Killing tensors and Killing vectors such that regular Killing tensors and Killing vectors form coefficients in the \mbox{$\beta$-expansion} of these generating functions.

Let us first define a $\beta$-dependent conformal Killing tensor as
\begin{equation}
	\boldsymbol{q}(\beta) = \bg + \beta^2\boldsymbol{Q} \,,
\end{equation}
where $\beta$ is a real parameter and $Q^{ab}=h^a{}_ch^{bc}$ is the first conformal Killing tensor. We also define a scalar function
\begin{equation}\label{eq:Abeta-def}
	A(\beta) = \sqrt{\frac{\det\boldsymbol{q}(\beta)}{\det\bg}} = \prod_{\nu} \left(1+\beta^2 x_{\nu}^2\right)\,.
\end{equation}
Using these definitions, we can introduce generating functions for Killing tensors and Killing vectors, respectively, in the form
\begin{equation}
	\bk(\beta) = A(\beta)\,\boldsymbol{q}^{-1}(\beta) \,,\qquad
	\bl(\beta) = \bk(\beta)\cdot\bxi \,.
\end{equation}
Here, $\bxi$ is a special Killing vector given by the divergence of ${\bh}$,
\begin{equation}
\label{xidef}
    \bxi = \frac{1}{2N-1} \,\boldsymbol{\nabla}\cdot\bh\;.
\end{equation}
The $\beta$-expansion of these functions can be written as
\begin{equation}
\label{eq:klbeta-expansion}
	\bk(\beta) = \sum_{k=0}^N \beta^{2k}\bk_{(k)} \,,\qquad
	\bl(\beta) = \sum_{k=0}^N \beta^{2k}\bl_{(k)} \,,
\end{equation}
thus generating Killing tensors $\bk_{(k)}$ and Killing vectors $\bl_{(k)}$. Similarly, the function $A(\beta)$ generates the polynomials $A^{(k)}$
\begin{equation}
\label{eq:Abeta-expansion}
	A(\beta) = \sum_{k=0}^N \beta^{2k}A^{(k)} \,.
\end{equation}
For a fixed parameter $\beta$, $\bk(\beta)$ is a linear combination of Killing tensors, therefore, $\bk(\beta)$ itself is a Killing tensor. Likewise, $\bl(\beta)$ is a Killing vector.

Apart from $\bl_{(k)}$, we also define alternative Killing vectors $\bs_{(\mu)}$, which are, up to normalization, the coordinate vectors,
\begin{equation}
\label{eq:s-def}
	\bs_{(\mu)} = \lambda a_{\mu}\framevec{\phi_{\mu}}
    = \sum_\nu \frac{J_{\nu}(a_{\mu}^2)}{\calU_{\mu}}\,\hatbeps_{\nu} \,.
\end{equation}

Since the unnormalized orthogonal frame of vectors \{$\beps_{\mu}$, $\hat{\beps}_{\mu}$\} \eqref{eq:OG-coframe} is suitable for performing the limit, let us provide explicit forms of Killing vectors and Killing tensors using this frame. For Killing vectors $\bl_{(k)}$ we have\footnote{Latin indices, unless explicitly indicated otherwise, go over the range $k,l,\ldots=0,\ldots,N-1$, and we do not use the Einstein summation convention for them.}
\begin{equation}
\label{eq:l-s-explicit}
	\bl_{(k)} = \sum_{\mu} \calA_{\mu}^{(k)}\bs_{(\mu)}  = \sum_{\mu} A_{\mu}^{(k)}\hatbeps_{\mu}\,.
\end{equation}
The Killing vector ${\bxi}$ reduces to
\begin{equation}
\label{eq:princip-kill-vect}
	\bxi = \bl_{(0)} = \sum_{\mu} \hatbeps_{\mu} = \sum_{\mu} \bs_{(\mu)} \,.
\end{equation}
The corresponding generating function reads
\begin{equation}
\label{eq:lbeta}
	\bl(\beta) = \sum_{\mu} \calA_{\mu}(\beta)\bs_{(\mu)} = \sum_{\mu} A_{\mu}(\beta)\hatbeps_{\mu} \,,
\end{equation}
where the functions $A_{\mu}(\beta)$ generate the polynomials $A_{\mu}^{(k)}$ and can be expressed in several ways
\begin{equation}
\label{eq:Amu-beta}
	A_{\mu}(\beta) = \sum_k \beta^{2k}A_{\mu}^{(k)} = \prod_{\overset{\scriptstyle{\nu}}{\scriptstyle{\nu\neq\mu}}} \left(1+\beta^2 x_{\nu}^2\right) = \frac{A(\beta)}{1+\beta^2 x_{\mu}^2} \,,
\end{equation}
with $A(\beta)$ given by \eqref{eq:Abeta-def}. Analogous expressions can be written for $\calA_{\mu}(\beta)$ and $\calA(\beta)$.

Motivated by the structure of Killing vectors, we also introduce a new set of Killing tensors $\br_{(\mu)}$,
\begin{equation}
\label{eq:r-def}
	\br_{(\mu)} = \sum_{\nu} \frac{J_{\nu}(a_{\mu}^2)}{\calU_{\mu}}\, \bpi_{\nu} \,,
\end{equation}
where $\bpi_{\mu}$ denote ``frame'' 2-tensors appearing explicitly in the metric, cf.~\eqref{eq:metric-frame},
\begin{equation}
\label{eq:pi-explicit}
	\bpi_{\mu} = \frac{X_{\mu}}{U_{\mu}}\,\beps_{\mu}\beps_{\mu} + \frac{U_{\mu}}{X_{\mu}}\,\hatbeps_{\mu}\hatbeps_{\mu} \,.
\end{equation}
Tensors $\br_{(\mu)}$ form a base of Killing tensors alternative to $\bk_{(k)}$ introduced above. Their equivalence can be observed from the relations analogous to \eqref{eq:l-s-explicit}
\begin{equation}
\label{eq:k-r-explicit}
	\bk_{(k)} = \sum_{\mu} \calA_{\mu}^{(k)}\br_{(\mu)} = \sum_{\mu} A_{\mu}^{(k)}\bpi_{\mu} \,.
\end{equation}
We see that ${\bk_{(k)}}$ are linear combinations with constant coefficients of tensors ${\br_{(\mu)}}$.

The generating function for Killing tensors ${\bk_{(k)}}$ adopts the form
\begin{equation}
\label{eq:kbeta}
	\bk(\beta)
    = \sum_{\mu} \calA_{\mu}(\beta)\,\br_{(\mu)}
    = \sum_{\mu} A_{\mu}(\beta)\,\bpi_{\mu}\,.
\end{equation}


\section{Equal-spin limit}
\label{sc:metric-limit}

In this section, we introduce an appropriate parametrization of the limit, which includes modifying the index notation used throughout Section \ref{sc:kerr-nut-ads}, and apply the limiting procedure to the general metric. We also study the limit case of a black hole using the Myers--Perry coordinate system. We will restrict our analysis to the case of vanishing NUT parameters and non-zero mass, i.e. $b_{\barmu}=0$, since it is the most relevant case from the physical point of view.

Let us emphasize that any limit of spacetime always strongly depends on the choice of the limiting procedure and used parametrization; see the classical work of Geroch \cite{Geroch:1969ca} and an illustration in, e.g., \cite{Bengtsson:2014fha}. One has to always carefully choose a limit interesting from the physical point of view. Different choices of the limiting procedure might focus on different aspects and thus would lead to different spacetimes after the limit. For example, one can zoom in on the regions near the black hole horizon by including a suitable rescaling during the limit, which would result in a near-horizon limit \cite{Lu2009,Chernyavsky2014,Xu2015,Galajinsky2016,Kolar2017}, or rescale asymptotic regions, which could reveal the asymptotic structure of spacetime. Our limiting procedure preserves (and possibly enhances) the symmetry structure of the spacetime. Moreover, all the outer regions of the black hole remain non-degenerate after the limit.

\begin{figure*}
\centering
\includegraphics[width=\textwidth]{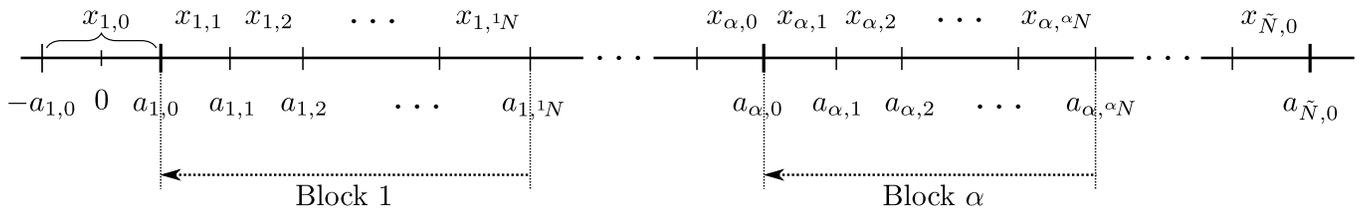}
\caption{New indexing and grouping of the rotational parameters \{$a_{\alpha,0}$, $a_{\alpha,\rho}$\} into blocks that have equal spin once the limit $a_{\alpha,\rho}\rightarrow a_{\alpha,0}$ has been performed. The coordinates \{$x_{\alpha,0}$, $x_{\alpha,\rho}$\} remain restricted by the rotational parameters as in \eqref{eq:new-x-ranges}.}
\label{fig:new-a-ordering}
\end{figure*}

\subsection{Preliminaries}
\label{sc:double-indexing}

In order to perform the limit, it will be convenient to modify the indexing of parameters and coordinates to reflect the structure that will emerge after the limit. Namely, assuming that the rotational parameters $a_{\mu}$ are ordered, we group them into $\tilN$ ``equal-spin'' blocks so that within each block all the rotations approach the same value. This means that instead of using a single Greek index $\mu$ (or $\nu$, $\kappa$, \dots), it will be more natural to use two Greek indices --- one from the beginning of the alphabet $\alpha$ ($\beta$, $\gamma$, \dots)\footnote{Strictly speaking, we should be using tilded indices $\tilde{\alpha}$ ($\tilde{\beta}$, $\tilde{\gamma}$, \dots) to label the blocks of equal rotations in order to clearly distinguish between indices and quantities before and after the limit. However, for the sake of simplicity and better readability, we will use tildes only over the names of the relevant quantities after the limit, e.g. $\tilx_{\alpha}$, $\tila_{\alpha}$.} to label the block of equal rotations and the second from the later parts of the alphabet $\rho$ ($\sigma$, $\tau$, \dots) to distinguish between the rotations inside the block.

The first rotation in a block is labeled as $a_{\alpha,0}$. It will remain unchanged after the limit and all the other rotations in the block, denoted by $a_{\alpha,\rho}$, will approach $a_{\alpha,0}$ as shown in Figure \ref{fig:new-a-ordering}. The indices go over the ranges
\begin{equation}
\begin{split}
	\alpha, \beta, \ldots & = 1, \ldots, \tilN \,, \\
	\rho, \sigma, \ldots & = 1, \ldots, \upal N \,.
\end{split}
\end{equation}
$\tilN$ is the number of blocks, and therefore the number of distinct rotational parameters remaining in the spacetime after the limit, while ${\upal N}$ is the number of parameters in the block $\alpha$ subject to the limit, and thus the number of additional rotations approaching the value $a_{\alpha,0}$. These numbers satisfy
\begin{equation}
	\tilN + \sum_{\alpha}\upal N = N \,.
\end{equation}

In the Lorentzian case we assume
\begin{equation}
	\upind{\tilN}N=0 \,.
\end{equation}
This means that the last block, which is related to the Wick-rotated radial coordinate, contains only a single parameter $a_{\tilN,0}$ not subject to the limit. This will allow us to obtain the Lorentzian metric using the Wick rotation in the same way as in \eqref{eq:Wick}.

All the other quantities such as coordinates and metric functions will be indexed in the same way.

The ranges of the coordinates \{$x_{\alpha,0}$, $x_{\alpha,\rho}$\} are
\begin{equation}
\label{eq:new-x-ranges}
\begin{split}
	a_{\alpha-1,\upind{\alpha-1}N} & < x_{\alpha,0} < a_{\alpha,0} \,,\\
	a_{\alpha,\rho-1} & < x_{\alpha,\rho} < a_{\alpha,\rho} \,,
\end{split}
\end{equation}
with the only exception being $x_{1,0}\in(-a_{1,0},a_{1,0})$.

When performing the limit $a_{\alpha,\rho}\rightarrow a_{\alpha,0}$, the rotational parameters $a_{\alpha,\rho}$ and the ranges of the coordinates $x_{\alpha,\rho}$ degenerate. Therefore, we will rescale them using the following parametrization
\begin{equation}\label{eq:parametrization}
\begin{aligned}
	a_{\alpha,0} &= \tila_{\alpha} \,, &
       x_{\alpha,0} &= \tilx_{\alpha} \,, &
       \phi_{\alpha,0} &= \tilphi_{\alpha} \,, \\
	a_{\alpha,\rho} &= \tila_{\alpha} + \upal a_{\rho}\,\varepsilon \,, &
       x_{\alpha,\rho} &= \tila_{\alpha} + \upal x_{\rho}\,\varepsilon \,, &
       \phi_{\alpha,\rho} &= \upal\phi_{\rho} \,,
\end{aligned}
\end{equation}
where $\varepsilon$ is a small limiting parameter, $\varepsilon\ll 1$. We have denoted the quantities that do not change in the limit using tildes and we will refer to the corresponding directions as \emph{primary coordinate directions}. We have also introduced new rescaled parameters $\upal a_{\rho}$ and coordinates $\upal x_{\rho}$, which remain well-defined after the limit, and we will refer to the corresponding directions as \emph{secondary coordinate directions}. Within one block, $\upal x_{\rho}$ are ordered~as
\begin{equation}
	0 < \upal x_1 < \upal a_1 < \upal x_2 < \upal a_2 < \ldots < \upal x_{\upal N} < \upal a_{\upal N} \,.
\end{equation}

As was mentioned earlier, the last block can be Wick-rotated to obtain the Lorentzian signature, cf.~\eqref{eq:Wick} and \eqref{eq:gauge},
\begin{equation}
\label{eq:Wick-limit}
	\tilx_{\tilN} = ir \,, \qquad \tilphi_{\tilN} = \lambda \tila_{\tilN} t \,, \qquad \tilde{b}_{\tilN} = iM \,,
\end{equation}
and
\begin{equation}
\label{eq:gauge-limit}
	\tila_{\tilN}^2 = -\frac{1}{\lambda} \,.
\end{equation}

\subsection{Limiting procedure}

Our goal is to perform the limit of the metric. For that we need to introduce two types of metric functions corresponding to the two sets of directions: tilded functions include only coordinates and parameters in the primary directions, and functions with an upper left index are constructed using only variables in the secondary directions. They are defined in a similar manner to the metric functions before the limit, only the sets of coordinates and parameters are modified, for example\footnote{We introduced a notation for sums (and products) of quantities in the secondary directions using an upper left index
\begin{equation*}
	\upal\sum_{\rho} \equiv \sum_{\rho=1}^{\upal N} \,.
\end{equation*}}
\begin{equation}
	\tilJ(\tila_{\alpha}^2) = \prod_{\beta} \left(\tilx_{\beta}^2 - \tila_{\alpha}^2\right) \,,\quad
	\upal J(\upal a_{\rho}) = \upal\prod_{\sigma} \left(\upal x_{\sigma} - \upal a_{\rho}\right) \,.
\end{equation}
An important difference is that the functions in the secondary directions, i.e.\ those including the rescaled variables $\upal x_{\rho}$ and $\upal a_{\rho}$, are not defined using squares, but only first powers.

Since we set $b_{\barmu}=0$, the on-shell metric functions after the limit are defined as
\begin{equation}
\label{eq:X-limit}
\begin{split}
	\tilX_{\baral} & = \lambda\tilcalJ(\tilx_{\baral}^2) \,,\\[0.25cm]
	\tilX_{\tilN} & = \lambda\tilcalJ(\tilx_{\tilN}^2)
	    - 2\tilb_{\tilN}\tilx_{\tilN} \prod_{\baral}
       \left(\tila_{\baral}^2-\tilx_{\tilN}^2\right)^{-\upind{\baral}N} \,,\\[0.1cm]
	\upind{\baral} X_{\rho} & =
       2\upind{\baral} x_{\rho} \upind{\baral}\calJ(\upind{\baral} x_{\rho}) \,,
\end{split}
\end{equation}
and $\upind{\tilN}X_{\rho}$ does not exist in the Lorentzian case since there is no ${\rho}$ in the last block. Barred indices are used in the same way as before, namely, to skip the Lorentzian sector, i.e. they go over the ranges
\begin{align}
	\baral, \barbe, \ldots & = 1, \ldots, \bartilN \,,\\[0.1cm]
    \bartilN & = \tilN-1 \,.
\end{align}

Now we can expand the unnormalized orthogonal frame of 1-forms \eqref{eq:OG-frame} in the limiting parameter ${\varepsilon}$
\begin{equation}
\begin{split}
	\beps^{\alpha,0} & \approx \tilbeps^{\alpha}\,, \qquad\,\,\,\,
	\hatbeps^{\alpha,0} \approx \htilbeps^{\alpha} \,, \\[0.2cm]
	\beps^{\alpha,\rho} & \approx \varepsilon\,\upal\beps^{\rho} \,, \qquad
	\hatbeps^{\alpha,\rho} \approx \frac{1}{\varepsilon} \frac{\tilJ(\tila_{\alpha}^2)}{2\lambda\tila_{\alpha}^2\tilcalU_{\alpha}} \, \upal\hatbeps^{\rho} \,.
\end{split}
\end{equation}
Here, $\approx$ denotes equality in the leading order in~$\varepsilon$. The 1-forms defined using the rescaled quantities on the right-hand sides of the equations read
\begin{equation}
\label{eq:ON-frame-limit}
\begin{split}
	\tilbeps^{\alpha} & = \bd \tilx_{\alpha} \,, \quad\,\,\,
	\htilbeps^{\alpha} = \sum_{\beta} \frac{\tilJ_{\alpha}(\tila_{\beta}^2)}{\lambda \tila_{\beta}\tilcalU_{\beta}} \tilPhi^{\beta} \,, \\[0.2cm]
	\upal\beps^{\rho} & = \bd \upal x_{\rho} \,, \quad
	\upal\hatbeps^{\rho} = \frac{\upal J_{\rho}(0)}{\upal\calJ(0)} \bd\tilphi_{\alpha} - \upal\sum_{\sigma} \frac{\upal J_{\rho}(\upal a_{\sigma})} {\upal a_{\sigma}\upal\calU_{\sigma}} \bd\upal\phi_{\sigma}  \,.
\end{split}
\end{equation}
We have written $\htilbeps^{\alpha}$ so that they have a similar form to $\hatbeps^{\mu}$ before the limit in \eqref{eq:OG-frame}, only instead of simple gradients $\bd\phi_{\mu}$ we had to introduce 1-forms $\tilPhi^{\alpha}$ given by
\begin{equation}
\label{eq:Phi-form}
	\tilPhi^{\alpha} = \frac{\upal J(0)}{\upal\calJ(0)} \bd\tilphi_{\alpha} - \upal\sum_{\rho} \frac{\upal J(\upal a_{\rho})} {\upal a_{\rho}\upal\calU_{\rho}} \bd\upal\phi_{\rho} \,.
\end{equation}
In Section \ref{sc:symmetries-limit}, we will show that they are K\"{a}hler potentials.

The dual unnormalized orthogonal frame of vectors \eqref{eq:OG-coframe} becomes
\begin{equation}
\begin{split}
	\beps_{\alpha,0} & \approx \tilbeps_{\alpha}\,, \qquad\quad
	\hatbeps_{\alpha,0} \approx \htilbeps_{\alpha} \,, \\[0.2cm]
	\beps_{\alpha,\rho} & \approx \frac{1}{\varepsilon}\upal\beps_{\rho} \,, \qquad
	\hatbeps_{\alpha,\rho} \approx \varepsilon \frac{2\lambda\tila_{\alpha}^2\tilcalU_{\alpha}}{\tilJ(\tila_{\alpha}^2)} \, \upal\hatbeps_{\rho} \,,
\end{split}
\end{equation}
where the frame vectors defined in terms of the rescaled quantities after the limit are
\begin{align}
	\tilbeps_{\alpha} & = \framevec{\tilx_{\alpha}} \,, &
	\htilbeps_{\alpha} &= \sum_{\beta} \frac{\tilcalJ_{\beta}(\tilx_{\alpha}^2)}{\tilU_{\alpha}} \tilPhi_{\beta} \,, \notag\\[-2ex]
	\upal\beps_{\rho} & = \framevec{\upal x_{\rho}} \,,     \label{eq:ON-coframe-limit}\\
 	&&&\mspace{-60mu}\upal\hatbeps_{\rho} = \frac{\upal\calJ(\upal x_{\rho})}{\upal U_{\rho}} \framevec{\tilphi_{\alpha}}
	- \upal\sum_{\sigma} \frac{\upal x_{\rho}\upal\calJ_{\sigma}(\upal x_{\rho})}{\upal U_{\rho}}  \framevec{\upal\phi_{\sigma}} \,.\notag
\end{align}
Here, the vectors $\tilPhi_{\alpha}$ read
\begin{equation}
\label{eq:Phi-vector}
	\tilPhi_{\alpha} = \lambda \tila_{\alpha} \left( \framevec{\tilphi_{\alpha}} + \upal\sum_{\rho} \framevec{\upal\phi_{\rho}} \right) \,.
\end{equation}
It turns out that they are Killing vectors, as will be shown in Section \ref{sc:symmetries-limit}. Notice that $\htilbeps_{\alpha}$ resemble $\hatbeps_{\mu}$ in \eqref{eq:OG-coframe}.

The orthogonal frame after the limit separates into two sets: the \emph{primary frame directions} \{$\tilbeps^{\alpha}$, \nolinebreak $\htilbeps^{\alpha}$\}, and the \emph{secondary frame directions} \{$\upal\beps^{\rho}$, \nolinebreak $\upal\hatbeps^{\rho}$\}\footnote{The index notation in the secondary frame directions is as follows. The left index (indicating which block a direction belongs to) is always placed at the top, whereas the position of the right index (distinguishing between the directions inside the block) reveals in a standard manner whether the concerned object is a form or a vector.}. This separation is valid only in the sense of tangent spaces since these directions do not correspond directly to the primary and the secondary coordinate directions --- the hatted 1-forms $\htilbeps^{\alpha}$ and $\upal\hatbeps^{\rho}$ contain angular coordinates in both primary and secondary coordinate directions. Moreover, primary and secondary frame directions are not integrable distributions of subspaces in the tangent spaces.

The splitting is respected by the duality relations between the frame of vectors and the frame of 1-forms
\begin{equation}
\label{eq:frame-duality-limit}
\begin{split}
	\tilbeps_{\alpha}\cdot\tilbeps^{\beta} & = \delta_{\alpha}^{\beta} \,, \qquad\qquad
	\htilbeps_{\alpha}\cdot\htilbeps^{\beta} = \delta_{\alpha}^{\beta} \,, \\[0.2cm]
	\upal\beps_{\rho}\cdot\upbe\beps^{\sigma} & = \delta_{\alpha\beta} \delta_{\rho}^{\sigma} \,, \qquad\,
	\upal\hatbeps_{\rho}\cdot\upbe\hatbeps^{\sigma} = \delta_{\alpha\beta} \delta_{\rho}^{\sigma} \,,
\end{split}	
\end{equation}
with all the other products being zero. The frames of the primary directions \{$\tilbeps^{\alpha}$, \nolinebreak $\htilbeps^{\alpha}$\} and in the individual blocks of the secondary directions \{$\upal\beps^{\rho}$, $\upal\hatbeps^{\rho}$\} become independent orthogonal frames.

The metric after the limit becomes
\begin{equation}
\label{eq:metric-limit}
	\bg \; \approx \; \tilbg \; - \; \sum_{\alpha} \frac{\tilJ(\tila_{\alpha}^2)}{2\lambda\tila_{\alpha}^2\tilcalU_{\alpha}} \, \upal\bg \,,
\end{equation}
where
\begin{align}
	\tilbg & = \sum_{\alpha} \left(\frac{\tilU_{\alpha}}{\tilX_{\alpha}} \tilbeps^{\alpha}\tilbeps^{\alpha} + \frac{\tilX_{\alpha}}{\tilU_{\alpha}}\htilbeps^{\alpha}\htilbeps^{\alpha} \right) \,, \label{eq:metric-limit-kerrnutads} \\[0.2cm]
	\upind{\baral}\bg & = \upind{\baral}\sum_{\rho} \left( \frac{\upind{\baral} U_{\rho}}{\upind{\baral} X_{\rho}} \, \upind{\baral}\beps^{\rho} \, \upind{\baral}\beps^{\rho} + \frac{\upind{\baral} X_{\rho}}{\upind{\baral} U_{\rho}} \, \upind{\baral}\hatbeps^{\rho} \, \upind{\baral}\hatbeps^{\rho} \right) \,,\quad
	\upind{\tilN}\bg = 0 \,. \label{eq:metric-limit-kahler}
\end{align}
The inverse metric in its limit form can be written as
\begin{equation}
\label{eq:metric-inv-limit}
	\bg^{-1} \; \approx \; \tilbg^{-1} \; - \; \sum_{\alpha} \frac{2\lambda\tila_{\alpha}^2\tilcalU_{\alpha}}{\tilJ(\tila_{\alpha}^2)} \, \upal\bg^{-1} \,,
\end{equation}
with
\begin{align}
	\tilbg^{-1} & = \sum_{\alpha} \left(\frac{\tilX_{\alpha}}{\tilU_{\alpha}} \tilbeps_{\alpha}\tilbeps_{\alpha} + \frac{\tilU_{\alpha}}{\tilX_{\alpha}}\htilbeps_{\alpha}\htilbeps_{\alpha} \right) \,,\qquad \label{eq:metric-inv-limit-kerrnutads}\\[0.2cm]
	\upind{\baral}\bg^{-1} & = \upind{\baral}\sum_{\rho} \left( \frac{\upind{\baral} X_{\rho}}{\upind{\baral} U_{\rho}} \, \upind{\baral}\beps_{\rho}\upind{\baral}\beps_{\rho} + \frac{\upind{\baral} U_{\rho}}{\upind{\baral} X_{\rho}} \, \upind{\baral}\hatbeps_{\rho}\upind{\baral}\hatbeps_{\rho} \right) \,,\quad
	\upind{\tilN}\bg^{-1} = 0 \,. \label{eq:metric-inv-limit-kahler}
\end{align}

Note, that tensors \eqref{eq:metric-inv-limit-kerrnutads} and \eqref{eq:metric-inv-limit-kahler} are individually inverse to the metrics \eqref{eq:metric-limit-kerrnutads} and \eqref{eq:metric-limit-kahler} on the respective subspaces spanned on the primary and the secondary frame directions, as can be seen using the duality relations \eqref{eq:frame-duality-limit}. Combined together, it gives that $\bg^{-1}$ is indeed the inverse of $\bg$.

More details of the limiting procedure applied to functions that appear in the metric are provided in Appendix~\ref{app:metric-funcs-limit}.

\subsection{Relation to generalized Kerr--NUT--(A)dS spacetimes}

It turns out that the metric \eqref{eq:metric-limit} is a special case of a more general metric described by Houri et al. in \cite{Houri2008}. In this paper, the authors study the \emph{generalized Kerr--NUT--(A)dS metric}, which possesses the principal tensor $\bh$ with both non-constant and constant eigenvalues, thus the tensor is not necessarily non-degenerate.

In our case, the principal tensor before the limit has non-constant and functionally independent eigenvalues $x_{\mu}$ \cite{Krtous2008}. However, after employing the limiting procedure, some of these eigenva\-lues become constant. Namely, all the eigenvalues $x_{\alpha,\rho}$ from the secondary blocks degenerate into the respective constant values $\tila_{\alpha}$ after the limit, while $\tilx_{\alpha}$ in the primary directions remain non-constant. Therefore, the geometry obtained by applying the equal-spin limiting procedure is indeed a subcase of the results published in \cite{Houri2008}. This also confirms the results of Oota and Yasui \cite{Oota2010}.

The number of non-constant and distinct constant non-zero eigenvalues is the same in our case and it corresponds to $\tilN$ in our notation. Our metrics $\upal\bg$ in the secondary blocks differ from the tensors that the authors of \cite{Houri2008} call \emph{K\"{a}hler metrics} only by a constant factor. Therefore, we will refer to $\upal\bg$ as K\"{a}hler metrics as well. The K\"{a}hler potentials identified in \cite{Houri2008} correspond to the 1-forms ${\tilPhi^{\alpha}}$ defined in \eqref{eq:Phi-form}.

Furthermore, the metric $\tilbg$ defined in \eqref{eq:metric-limit-kerrnutads} has the form of the Kerr--NUT--(A)dS metric analogous to the metric of the entire spacetime \eqref{eq:metric-frame} but for the primary directions only. Thus we shall refer to it as the \emph{Kerr--NUT--(A)dS part}.

However, we should emphasize that $\upal\bg$ represent the K\"{a}hler metrics only formally as there seems to be no decomposition of the original Kerr--NUT--(A)dS manifold into a direct product of the Kerr--NUT--(A)dS part and the K\"{a}hler manifolds. As was already mentioned, the primary and the secondary directions form only subspaces of tangent spaces. Metrics $\upal\bg$ act on these secondary subspaces of the tangent space, but these subspaces are not integrable to form independent K\"ahler manifolds.

\subsection{Black hole}

When applying the limiting procedure to the Jacobi transformation that introduces \nopagebreak the Myers--Perry coordinates, it turns out that we can define analogous transformations for the primary and the secondary directions separately. Indeed, the limit form of \eqref{eq:jacobi-mu} reads
\begin{equation}
\label{eq:jacobi-mu-limit}
	\mu_{\baral,0}^2 \approx \tilmu_{\baral}^2 \,\, \upind{\baral}\mu_0^2 \,,\quad
	\mu_{\baral,\rho}^2 \approx \tilmu_{\baral}^2 \,\, \upind{\baral}\mu_{\rho}^2 \,,\quad
	\mu_0^2 \approx \tilmu_0^2 \,,
\end{equation}
where we have denoted\footnote{Functions decorated with both a bar and a tilde are defined analogously to barred functions in Section \ref{sc:kerr-nut-ads}, i.e. without $\tilx_{\tilN}$ and $\tila_{\tilN}$.}
\begin{equation}
\label{eq:jacobi-mu-limit-def}
\begin{aligned}
	\tilmu_{\baral}^2 & = \frac{\bartilJ(\tila_{\baral}^2)}{-\tila_{\baral}^2\btilcalU_{\baral}} \,,&
	\tilmu_0^2 &= \frac{\bartilJ(0)}{\btilcalJ(0)} = \frac{\bartilA^{(\bartilN)}}{\btilcalA^{(\bartilN)}} \,, \\[0.0cm]
	\upind{\baral}\mu_{\rho}^2 & = \frac{\upind{\baral}J(\upind{\baral}a_{\rho})}{-\upind{\baral}a_{\rho} \upind{\baral}\calU_{\rho}} \,,\quad&
	\upind{\baral}\mu_0^2 &= \frac{\upind{\baral}J(0)}{\upind{\baral}\calJ(0)} = \frac{\upind{\baral}A^{(\upind{\baral}N)}}{\upind{\baral}\calA^{(\upind{\baral}N)}} \,.
\end{aligned}
\end{equation}
The advantage of defining the two sets of Myers--Perry coordinates in this way is that a constraint similar to \eqref{eq:sphere-mu} holds for the primary and all secondary sets independently
\begin{equation}
\label{eq:sphere-limit-mu}
	\sum_{\baral=0}^{\bartilN}\tilmu_{\baral}^2 = 1 \,,\qquad
	\sum_{\rho=0}^{\upind{\baral}N}\upind{\baral}\mu_{\rho}^2 = 1 \,.
\end{equation}

The metric after the limit in these coordinates reads
\begin{widetext}
\vspace{-2ex}
\begin{equation}
\label{eq:metric-mu-limit}
\begin{split}
	\bg \approx & -\left(1-\lambda\tilde{R}^2\right) \bd t^2
	+ \frac{2Mr}{\tilSig} \prod_{\barbe}\left(r^2+\tila_{\barbe}^2\right)^{-\upind{\barbe}N}
	\left[\bd t + \sum_{\baral} \frac{\tila_{\baral}\tilmu_{\baral}^2}{1+\lambda\tila_{\baral}^2} \left(\tilPhi^{\baral}-\lambda\tila_{\baral}\bd t\right)\right]^2
  \\[0ex]
	& + \frac{\tilSig}{\tilDel_r}\bd r^2 + r^2\bd\tilmu_0^2 + \sum_{\baral} \frac{r^2+\tila_{\baral}^2}{1+\lambda\tila_{\baral}^2} \left( \bd\tilmu_{\baral}^2 + \tilmu_{\baral}^2 \, \upind{\baral}\bg_{_{\textrm{Eucl}}} \right)
	+ \frac{\lambda}{1-\lambda\tilde{R}^2} \left( r^2\tilmu_0\bd\tilmu_0 + \sum_{\baral} \frac{r^2+\tila_{\baral}^2}{1+\lambda\tila_{\baral}^2} \tilmu_{\baral}\bd\tilmu_{\baral} \right)^2 \,,
\end{split}
\end{equation}
\end{widetext}
where $\upind{\baral}\bg_{_{\textrm{Eucl}}}$ denote ($2\,\upind{\baral}N{+}2$)-dimensional Euclidean metrics in the multi-polar coordinates
\begin{equation}
\label{eq:metric-flat-mu-limit}
	\upind{\baral}\bg_{_{\textrm{Eucl}}}\!\! = \bd\upind{\baral}\mu_0^2 + \upind{\baral}\mu_0^2\bd\tilphi_{\baral}^2 + \upind{\baral}\sum_{\rho}\! \left( \bd\upind{\baral}\mu_{\rho}^2 + \upind{\baral}\mu_{\rho}^2\bd\upind{\baral}\phi_{\rho}^2 \right) ,
\end{equation}
and
\begin{equation}
\label{eq:bh-func-mu-limit}
\begin{split}
	& 1 - \lambda\tilde{R}^2 = \left(1-\lambda r^2\right) \left(\tilmu_0^2 + \sum_{\baral}\frac{\tilmu_{\baral}^2}{1+\lambda\tila_{\baral}^2}\right) \,, \\[0.5ex]
	& \tilDel_r = \left(1-\lambda r^2\right) \prod_{\baral} \left(r^2+\tila_{\baral}^2\right) - 2Mr \prod_{\baral}\left(r^2+\tila_{\baral}^2\right)^{-\upind{\baral}N} \,,\\[0ex]
	& \tilSig = \left(\tilmu_0^2 + \sum_{\baral} \frac{r^2\tilmu_{\baral}^2}{r^2 + \tila_{\baral}^2}\right) \prod_{\barbe}\left(r^2 + \tila_{\barbe}^2\right) \,.
\end{split}\raisetag{10ex}
\end{equation}\\[-2ex]
Notice that these expressions are similar to \eqref{eq:bh-func-mu} before the limit --- except for the second term of $\tilDel_r$, which is multiplied by an additional factor emerging from applying the limit on quantities in the secondary directions. The 1-forms $\tilPhi^{\baral}$ in these coordinates read
\begin{equation}
	\tilPhi^{\baral} = \upind{\baral}\mu_0^2\,\bd\tilphi_{\baral}
    + \upind{\baral}\sum_{\rho} \upind{\baral}\mu_{\rho}^2\,\bd\upind{\baral}\phi_{\rho} \,.
\end{equation}

As one can see, the secondary blocks become spherically symmetric after the limit. In particular, they can be viewed as ($2\,\upind{\baral}N{+}1$)\nolinebreak-dimensional spheres given by the constraints \eqref{eq:sphere-limit-mu}, embedded in ($2\,\upind{\baral}N{+}2$)\nolinebreak-dimensional flat spaces, described by the metrics $\upind{\baral}\bg_{_{\textrm{Eucl}}}$ \eqref{eq:metric-flat-mu-limit}. Moreover, in the full spacetime metric \eqref{eq:metric-mu-limit}, each sphere is coupled solely to the coordinate $\tilmu_{\baral}$ in the corresponding primary direction. The full metric after the limit thus has a similar form to the metric \eqref{eq:metric-mu} before the limit, only 2-forms $\bd\phi_{\barnu}^2$ have been replaced with the spheres~$\upind{\baral}\bg_{_{\textrm{Eucl}}}$. The only other occurrence of the secondary directions is in the 1-forms $\tilPhi^{\baral}$, which play the role of K\"{a}hler potentials as is discussed in the next section.

Notice that the metric in the Myers--Perry coordinates no longer clearly separates into the Kerr--NUT--(A)dS part $\tilbg$ and the K\"{a}hler metrics $\upal\bg$, but they are rather combined together.


\section{Reconstructing original symmetries}
\label{sc:symmetries-limit}

This section focuses on explicit and hidden symmetries of the resulting spacetime after the limit $a_{\alpha,\rho}\rightarrow a_{\alpha,0}$ has been performed. We reconstruct the original number of Killing vectors and Killing tensors, thus showing that the symmetry group has not been reduced during the limiting procedure. We expect the symmetry group to be enhanced, however, this will be shown explicitly only in the case of six dimensions, which is discussed in Section~\ref{sc:examples}.

\subsection{Principal tensor}

Let us first discuss the limit form of the principal tensor. Applying the limiting procedure as before, the principal tensor \eqref{eq:h-explicit} becomes
\begin{equation}
\label{eq:h-limit}
	\bh \approx \sum_{\alpha} \left[ \tilx_{\alpha}\tilbeps^{\alpha}\wedge\htilbeps^{\alpha} + \frac{\tilJ(\tila_{\alpha}^2)}{2\lambda\tila_{\alpha}\tilcalU_{\alpha}} \bomega^{\alpha} \right] \,,
\end{equation}
where $\bomega^{\alpha}$ are defined as
\begin{equation}
	\bomega^{\baral} = \upind{\baral}\sum_{\rho} \upind{\baral}\beps^{\rho} \wedge \upind{\baral}\hatbeps^{\rho} \,,\qquad
	\bomega^{\tilN} = 0 \,.
\end{equation}

Similarly to the metric, we have obtained the principal tensor in the form that represents a special case of the results published in \cite{Houri2008}. Our tensors $\bomega^{\alpha}$ differ from the tensors that the authors of \cite{Houri2008} call the K\"{a}hler forms corresponding to the K\"{a}hler metrics only by a constant factor. Therefore, $\bomega^{\alpha}$ formally represent the K\"{a}hler forms.

According to its general definition, a K\"{a}hler form is closed. It can be shown that $\bd\bomega^{\alpha}=0$, thus in our case this requirement is indeed satisfied. Moreover, there exists a K\"{a}hler potential, which is in our case represented by $\tilPhi^{\alpha}$ defined earlier in \eqref{eq:Phi-form}. Namely, the following equality holds
\begin{equation}
	\bomega^{\alpha} = \bd\tilPhi^{\alpha} \,.
\end{equation}

\subsection{Killing vectors}

A straightforward way to obtain the limit of Killing vectors is to focus on the Killing vectors associated with the coordinates $\phi_\mu$. The limit of their rescaled version $\bs_{(\mu)}$, in the new indexing scheme, is
\begin{equation}
	\bs_{(\alpha,0)} \approx \tils_{(\alpha)} \,,\qquad
	\bs_{(\alpha,\rho)} \approx \upal \bs_{(\rho)} \,,
\end{equation}
with $\tils_{(\alpha)}$ and $\upal \bs_{(\rho)}$ defined similarly to \eqref{eq:s-def},
\begin{equation}
	\tils_{(\alpha)} = \lambda\tila_{\alpha}\framevec{\tilphi_{\alpha}} \,,\qquad
	\upal \bs_{(\rho)} = \lambda\tila_{\alpha}\framevec{\upal\phi_{\rho}} \,.
\end{equation}
We have thus obtained $\tilN$ Killing vectors $\tils_{(\alpha)}$ in the primary directions and $\sum_{\alpha}\upal N$ Killing vectors $\upal\bs_{(\rho)}$ in the secondary directions, which gives in total $N$ explicit symmetries --- the same number as before the limit.

The Killing vectors $\tilPhi_{\alpha}$ introduced in \eqref{eq:Phi-vector} can now be rewritten as
\begin{equation}
\label{eq:Phi-s}
	\tilPhi_{\alpha} = \tils_{(\alpha)} + \upal\sum_{\rho}\upal \bs_{(\rho)} \,.
\end{equation}

Inspired by the first equality in \eqref{eq:l-s-explicit}, we can also introduce Killing vectors $\till_{(r)}$ and $\upal\bl_{(p)}$ as linear combinations of the coordinate Killing vectors $\tils_{(\alpha)}$ and $\upal\bs_{(\rho)}$ in the Kerr--NUT--(A)dS part $\tilbg$ and the K\"{a}hler parts $\upal\bg$ of the metric, respectively
\begin{equation}
\begin{split}
	\till_{(r)} & = \sum_{\alpha} \tilcalA_{\alpha}^{(r)} \tils_{(\alpha)} \,,\\[0.2cm]
	\upal\bl_{(p)} & = \upal\sum_{\rho} \upal\calA_{\rho}^{(p)} \, \upal\bs_{(\rho)} = \upal\sum_{\rho} \upal A_{\rho}^{(p)} \, \upal\hatbeps_{\rho} \,.
\end{split}
\end{equation}
Here, the indices $r$ and $p$ go over the ranges
\begin{eqnarray}
	r & = 0, \ldots, \tilN-1 \,, \\[0.1cm]
	p & = 0, \ldots, \upal N-1 \,,
\end{eqnarray}
and the functions $\tilcalA_{\alpha}^{(r)}$, $\upal\calA_{\rho}^{(p)}$ and $\upal A_{\rho}^{(p)}$ are as in \eqref{eq:app-A-alpharho}.
Similarly to \eqref{eq:princip-kill-vect}, we also introduce the special Killing vectors $\tilxi$ and $\upal\bxi$
\begin{equation}
\label{eq:limitxi}
\begin{aligned}
	\tilxi & = \till_{(0)} = \sum_{\alpha} \tils_{(\alpha)} \,,\\
	\upal\bxi & = \upal\bl_{(0)} = \upal\sum_{\rho} \upal\bs_{(\rho)} = \upal\sum_{\rho} \upal\hatbeps_{\rho} \,.
\end{aligned}
\end{equation}

Notice, that we have not listed relations between $\till_{(r)}$ and $\htilbeps_{\alpha}$ similar to the second equality in \eqref{eq:l-s-explicit}. Indeed, such relations do not exist. It turns out that the Killing vectors $\till_{(r)}$ and $\tils_{(\alpha)}$ need to be ``improved'' in order to fulfill such relations.

To show this, let us approach the limit from a different direction. We perform the limit of the generating function for Killing vectors \eqref{eq:lbeta}. Using an expression analogous to \eqref{eq:Amu-beta} for $\calA_{\mu}(\beta)$ and \eqref{eq:Phi-s}, we obtain
\begin{equation}
\label{eq:lbeta-limit}
	\bl(\beta) \approx \prod_{\gamma} \left(1+\beta^2\tila_{\gamma}^2\right)^{\upga N} \tilL(\beta) \,,
\end{equation}
with
\begin{equation}
\label{eq:tilLbeta-tilPhi}
	\tilL(\beta) = \sum_{\alpha} \tilcalA_{\alpha}(\beta) \tilPhi_{\alpha}\,.
\end{equation}
Realizing that the multiplicative prefactor ${\prod_{\gamma}(1+\beta^2\tila_{\gamma}^2)^{\upga N}}$ in \eqref{eq:lbeta-limit} is constant on the manifold, we can understand $\tilL(\beta)$ also as a generating function for Killing vectors. Namely, it will be the generating function for new Killing vectors $\tilL_{(r)}$,
\begin{equation}
\label{eq:tilLbeta-tilLr}
	\tilL(\beta) = \sum_r \beta^{2r}\tilL_{(r)} \,.
\end{equation}
Comparing \eqref{eq:tilLbeta-tilLr} with \eqref{eq:tilLbeta-tilPhi} and using the relation \eqref{eq:ON-coframe-limit} between $\tilPhi_\alpha$ and $\htilbeps_{\alpha}$, we find $\tilL_{(r)}$ to be fully analogous to \eqref{eq:l-s-explicit}
\begin{equation}
\label{eq:lbetaPhi}
	\tilL_{(r)} = \sum_{\alpha} \tilcalA_{\alpha}^{(r)} \tilPhi_{\alpha}
                = \sum_{\alpha} \tilA_{\alpha}^{(r)} \htilbeps_{\alpha} \,.
\end{equation}
The special Killing vector $\tilXi$ associated with $\tilL_{(r)}$ reads
\begin{equation}
\label{eq:Xi-eps}
	\tilXi = \tilL_{(0)} = \sum_{\alpha} \tilPhi_{\alpha}  = \sum_{\alpha} \htilbeps_{\alpha} \,.
\end{equation}

The generating function $\tilL(\beta)$ thus provides us with the Killing vectors $\tilL_{(r)}$ or $\tilPhi_{\alpha}$, which ``improve'' the Killing vectors $\till_{(r)}$ and $\tils_{(\alpha)}$ in such a way that \eqref{eq:lbetaPhi} holds. Of course, any of these sets of Killing vectors can be used since they carry the same information. Indeed, the ``improved'' Killing vectors are related to the previously defined Killing vectors as
\begin{align}
\label{eq:L-lxi}
	\tilL_{(r)} &= \till_{(r)} + \sum_{\alpha} \tilcalA_{\alpha}^{(r)} \, \upal\bxi \,,\\[-0.5ex]
\label{eq:Phi-sxi}
	\tilPhi_{\alpha} &= \tils_{(\alpha)} + \upal\bxi
\end{align}
and
\begin{equation}\label{eq:Xi-xi}
    \tilXi = \tilxi + \sum_{\alpha} \upal\bxi \,.
\end{equation}

\subsection{Killing tensors}

To calculate the limit of Killing tensors, we shall proceed in a similar manner to Killing vectors and apply the limiting procedure to the Killing tensors $\br_{(\mu)}$, cf.~\eqref{eq:r-def}.

As a preliminary, we consider the limit of $\bpi_{\mu}$ defined in \eqref{eq:pi-explicit}
\begin{equation}
\label{eq:pi-limit}
	\bpi_{\alpha,0} \approx \tilpi_{\alpha} \,,\qquad
	\bpi_{\alpha,\rho} \approx -\frac{2\lambda\tila_{\alpha}^2\tilcalU_{\alpha}}{\tilJ(\tila_{\alpha}^2)} \, \upal\bpi_{\rho} \,.
\end{equation}
Here, the tensors $\tilpi_{\alpha}$ and $\upal\bpi_{\rho}$ are
\begin{equation}
\begin{aligned}
	\tilpi_{\alpha} & = \frac{\tilX_{\alpha}}{\tilU_{\alpha}} \tilbeps_{\alpha}\tilbeps_{\alpha} + \frac{\tilU_{\alpha}}{\tilX_{\alpha}}\htilbeps_{\alpha}\htilbeps_{\alpha} \,,\\[0.2cm]
	\upal\bpi_{\rho} & = \frac{\upal X_{\rho}}{\upal U_{\rho}} \, \upal\beps_{\rho}\upal\beps_{\rho} + \frac{\upal U_{\rho}}{\upal X_{\rho}} \, \upal\hatbeps_{\rho}\upal\hatbeps_{\rho} \,.
\end{aligned}
\end{equation}
We also define auxiliary tensors $\tilk_{(r)}$ and $\tilr_{(\alpha)}$ using relations analogous to \eqref{eq:k-r-explicit},
\begin{equation}\label{eq:tilk-tilr-tilpi}
	\tilk_{(r)} = \sum_{\alpha} \tilcalA_{\alpha}^{(r)} \tilr_{(\alpha)} = \sum_{\alpha} \tilA_{\alpha}^{(r)} \tilpi_{\alpha} \,,
\end{equation}
with the corollary
\begin{equation}
	\tilk_{(0)} = \tilbg^{-1} \,.
\end{equation}
It should be emphasized that the tensors $\tilk_{(r)}$ and $\tilr_{(\alpha)}$ \emph{are not}, in general, Killing tensors.

The leading term in the expansion of the Killing tensors $\br_{(\mu)}$ now becomes
\begin{equation}
\label{eq:r-limit}
	\varepsilon\br_{(\alpha,0)} \approx -\upal\sum_{\rho} \frac{\lambda\tila_{\alpha}}{\upal a_{\rho}} \, \upal\br_{(\rho)} \,,\qquad
	\varepsilon\br_{(\alpha,\rho)} \approx \frac{\lambda\tila_{\alpha}}{\upal a_{\rho}} \, \upal\br_{(\rho)} \,,
\end{equation}
where $\upal\br_{(\rho)}$,
\begin{equation}
	\upal\br_{(\rho)} = \upal\sum_{\sigma} \frac{\upal J_{\sigma}(\upal a_{\rho})}{\upal\calU_{\rho}} \, \upal\bpi_{\sigma} \,,
\end{equation}
are independent Killing tensors after the limit associated with the secondary directions. Again, in analogy with \eqref{eq:k-r-explicit} we also introduce Killing tensors $\upal\bk_{(p)}$ as
\begin{equation}
\label{eq:k-secondary}
	\upal\bk_{(p)}
          = \upal\sum_{\rho} \upal\calA_{\rho}^{(p)} \, \upal\br_{(\rho)}
          = \upal\sum_{\rho} \upal A_{\rho}^{(p)} \, \upal\bpi_{\rho} \,.
\end{equation}
In this case, both $\upal\bk_{(p)}$ and $\upal\br_{(\rho)}$ \emph{are} Killing tensors: $\upal\br_{(\rho)}$ have been obtained as a limit of Killing tensors and $\upal\bk_{(p)}$ are just linear combinations of $\upal\br_{(\rho)}$ with constant coefficients.

The Killing tensors $\upal\bk_{(p)}$ and $\upal\br_{(\rho)}$ are directly related to the K\"{a}hler parts of the metric $\upal\bg$. For $p=0$ we even have
\begin{equation}
\label{eq:k0=kahler}
	\upal\bk_{(0)} = \upal\bg^{-1} \,,
\end{equation}
thus the K\"{a}hler metrics are Killing tensors as well.

Inspecting the first expansion in \eqref{eq:r-limit}, we see that the limiting procedure for $\br_{(\mu)}$ extracts only the Killing tensors $\upal\br_{(\rho)}$ in the secondary directions. It does not provide any additional information about the primary directions. Therefore, we need a different approach to obtain Killing tensors related to the primary directions.

Similarly to Killing vectors, let us apply the limiting procedure to the generating function $\bk(\beta)$ for Killing tensors \eqref{eq:kbeta}. Using \eqref{eq:Amu-beta} and the limit form of $\bpi_{\mu}$ \eqref{eq:pi-limit}, we obtain
\begin{equation}
\label{eq:kbeta-limit}
	\bk(\beta) \approx \prod_{\gamma} \left(1+\beta^2\tila_{\gamma}^2\right)^{\upga N} \tilK(\beta) \,.
\end{equation}
Since $\bk(\beta)$ and $\tilK(\beta)$ differ only by a constant factor, $\tilK(\beta)$ also generates Killing tensors. We use this generating function to generate Killing tensors associated with the primary directions after the limit.

Using \eqref{eq:k0=kahler}, $\tilK(\beta)$ can be written in the form
\begin{equation}
	\tilK(\beta) = \sum_r \beta^{2r}\tilK_{(r)} - \beta^{2\tilN} \sum_{\alpha} \frac{2\lambda\tila_{\alpha}^2\tilcalU_{\alpha}}{1+\beta^2\tila_{\alpha}^2} \, \upal\bg^{-1} \,,
\end{equation}
where we also separated the first $\tilN$ powers of $\beta^2$ from the remaining ones.
Because of the higher powers of $\beta^2$ in the second sum, $\tilK(\beta)$ is no longer a direct analogue of its before-the-limit counterpart $\bk(\beta)$ in \eqref{eq:klbeta-expansion}. Nevertheless, extra Killing tensors corresponding to these higher powers of $\beta^2$ are rather trivial --- they are just linear combinations of the Killing tensors $\upal\bg^{-1}$ with constant coefficients. The more interesting part of $\tilK(\beta)$ is given by the first $\tilN$ powers of $\beta^2$ contained in the first sum. The coefficients define new Killing tensors $\tilK_{(r)}$, which are given by
\begin{equation}
\label{eq:K-limit}
	\tilK_{(r)} = \tilk_{(r)} - \sum_{\alpha} \tilB^{(r)}(\tila_{\alpha}^2) \frac{2\lambda\tila_{\alpha}^2\tilcalU_{\alpha}}{\tilJ(\tila_{\alpha}^2)} \, \upal\bg^{-1} \,,
\end{equation}
with functions $\tilB^{(r)}$ defined as
\begin{equation}
	\tilB^{(r)}(\tila^2) = \sum_{n=0}^r \tilA^{(n)}(-\tila^2)^{r-n} \,.
\end{equation}
Since these are non-trivial functions on spacetime, the linear combination of the Killing tensors $\upal\bg^{-1}$ in \eqref{eq:K-limit} does not have constant coefficients and, therefore, $\tilk_{(r)}$ cannot be expected to be Killing tensors.

For $r=0$ we get
\begin{equation}
	\tilK_{(0)} \approx \bg^{-1} \,.
\end{equation}

Finally, we can introduce another set of Killing tensors $\tilR_{(\alpha)}$, again using a formula analogous to \eqref{eq:k-r-explicit}
\begin{equation}
\label{eq:K-R-limit}
	\tilK_{(r)} = \sum_{\alpha} \tilcalA_{\alpha}^{(r)} \tilR_{(\alpha)} \,.
\end{equation}
With the help of an expression for $\tilcalA_{\alpha}^{(r)}$ analogous to \eqref{eq:id-A-sum-k}, \eqref{eq:K-limit} and \eqref{eq:tilk-tilr-tilpi} as well as some algebra, it can be shown that $\tilR_{(\alpha)}$ read
\begin{equation}
\label{eq:R-limit}
\begin{split}
	\tilR_{(\alpha)} = \tilr_{(\alpha)} & + \sum_{\overset{\scriptstyle{\beta}}{\scriptstyle{\beta\neq\alpha}}}
	\frac{2\lambda\tila_{\beta}^2}{\tila_{\alpha}^2-\tila_{\beta}^2} \frac{\tilcalU_{\beta}}{\tilcalU_{\alpha}} \frac{\tilJ(\tila_{\alpha}^2)-\tilJ(\tila_{\beta}^2)}{\tilJ(\tila_{\beta}^2)} \, \upbe\bg^{-1} \\[0.2cm]
	& - 2\lambda\tila_{\alpha}^2 \sum_{\beta}\left(\tilx_{\beta}^2-\tila_{\alpha}^2\right)^{-1} \upal\bg^{-1} \,.
\end{split}\raisetag{10ex}
\end{equation}

To summarize, we have obtained $\tilN$ Killing tensors $\tilK_{(r)}$ in the primary directions and $\sum_{\alpha}\upal N$ Killing tensors $\upal\bk_{(p)}$ in the secondary directions, thus reconstructing the original number of $N$ hidden symmetries. Alternatively, we can use equivalent sets of Killing tensors $\tilR_{(\alpha)}$ in the primary directions and  $\upal\br_{(\rho)}$ in the secondary directions.


\section{Equally-spinning black holes}
\label{sc:examples}

This section includes two explicit examples of the general results obtained in Section \ref{sc:metric-limit} and \ref{sc:symmetries-limit} that are interesting from the physical point of view. Namely, we consider metrics with the Lorentzian signature describing higher-dimensional black holes, and we set all their rotational parameters equal. We present examples of black holes in $D=2N$ and in six dimensions.

\subsection{$\boldsymbol{D=2N}$}

Setting all the rotational parameters equal in the Lorentzian case means that there are only two equal-spin blocks $\alpha\in\{1,\tilN\}$, with the latter being reserved for the Wick rotation. Therefore, only the secondary directions within the first block are subject to the limiting procedure, and their number is $\upind{1}N=N-2$. The limit in this case is characterized by $a_{1,\rho}\rightarrow a_{1,0}$, where the index labeling the secondary directions acquires the values
\begin{equation}
	\rho = 1, \ldots, N-2 \,,
\end{equation}
unless indicated otherwise. The parametrization \eqref{eq:parametrization} thus adopts the form
\begin{equation}
\label{eq:D=2N-parametrization}
\begin{array}{lll}
	a_{1,0} = \tila_1 \,, & x_{1,0} = \tilx_1 \,, & \phi_{1,0} = \tilphi_1 \,, \\[0.1cm]
	a_{1,\rho} = \tila_1 + \upind{1}a_{\rho}\,\varepsilon \,, \hphantom{a} & x_{1,\rho} = \tila_1 + \upind{1}x_{\rho}\,\varepsilon \,, \hphantom{a} & \phi_{1,\rho} = \upind{1}\phi_{\rho} \,,\\[0.1cm]
	a_{\tilN,0} = \tila_{\tilN} \,, & x_{\tilN,0} = \tilx_{\tilN} \,, & \phi_{\tilN,0} = \tilphi_{\tilN} \,.
\end{array}
\end{equation}

We shall proceed from the generally spinning metric after the limit and use its form \eqref{eq:metric-mu-limit} in the Myers--Perry coordinates \{$t$, $r$, $\tilmu_0$, $\tilmu_1$, $\tilphi_1$, $\upind{1}\mu_0$, $\upind{1}\mu_{\rho}$, $\upind{1}\phi_{\rho}$\}. Since the equally-spinning black hole retains only a single rotational parameter after the limit, let us write
\begin{equation}
\label{eq:D=2N-rename-a}
	\tila_1 \equiv \tila \,.
\end{equation}
Moreover, the second limiting block is related just to the Lorentzian sector and it is renamed in the Wick rotation \eqref{eq:Wick-limit} and by imposing the gauge condition \eqref{eq:gauge-limit}. Thus only a single primary direction and the corresponding block of secondary directions remain. Therefore, we shall further simplify the notation by dropping the index ``1'' labeling the only limiting block
\begin{align}
	\tilphi_1 & \equiv \tilphi \,,\hspace{1cm}
	\upind{1}\phi_{\rho} \equiv \phi_{\rho} \,,\label{eq:D=2N-rename-phi} \\[0.2cm]
	\upind{1}\mu_0 & \equiv \mu_0 \,,\hspace{0.85cm}
	\upind{1}\mu_{\rho} \equiv \mu_{\rho} \,.
\end{align}

As in the general case, the coordinates \{$\tilmu_0$, $\tilmu_1$\} in the primary directions and \{$\mu_0$, $\mu_{\rho}$\} in the secondary directions are not independent --- they are constrained by the conditions
\begin{equation}
\label{eq:D=2N-sphere}
	\tilmu_0^2 + \tilmu_1^2 = 1 \,,\qquad
	\sum_{\rho=0}^{N-2} \mu_{\rho}^2 = 1 \,.
\end{equation}
The metric of an equally-spinning black hole then becomes
\begin{equation}
\label{eq:D=2N-metric}
\begin{split}
	\bg \approx & -\left(1-\lambda\tilde{R}^2\right) \bd t^2 \\[0.2cm]
	& + \frac{2Mr}{\tilSig} \left(r^2+\tila^2\right)^{2-N} \left[\bd t + \frac{\tila\tilmu_1^2}{1+\lambda \tila^2} \left(\tilPhi-\lambda \tila\bd t\right)\right]^2 \\[0.2cm]
	& + \frac{\tilSig}{\tilDel_r}\bd r^2 + r^2\bd\tilmu_0^2 + \frac{r^2+\tila^2}{1+\lambda \tila^2} \left( \bd\tilmu_1^2 + \tilmu_1^2\, \bg_{_{\textrm{Eucl}}} \right) \\[0.2cm]
	& + \frac{\lambda}{1-\lambda\tilde{R}^2} \left( r^2\tilmu_0\bd\tilmu_0 + \frac{r^2+\tila^2}{1+\lambda \tila^2} \tilmu_1\bd\tilmu_1 \right)^2 \,,
\end{split}
\end{equation}
where
\begin{equation}
	\tilPhi \equiv \tilPhi^1 = \mu_0^2 \bd\tilphi + \sum_{\rho} \mu_{\rho}^2\bd\phi_{\rho} \,,
\end{equation}
and $\bg_{_{\textrm{Eucl}}}$ denotes a ($2N{-}2$)-dimensional Euclidean metric in the multi-polar coordinates
\begin{equation}
\label{eq:D=2N-metric-flat}
	\bg_{_{\textrm{Eucl}}} = \bd\mu_0^2 + \mu_0^2 \bd\tilphi^2 + \sum_{\rho} \left( \bd\mu_{\rho}^2 + \mu_{\rho}^2\bd\phi_{\rho}^2 \right) \,.
\end{equation}
The metric functions adopt the form
\begin{equation}
\begin{split}
	1-\lambda\tilde{R}^2 & = \left(1-\lambda r^2\right) \left(\tilmu_0^2 + \frac{\tilmu_1^2}{1+\lambda \tila^2}\right) \,,\\[0.2cm]
	\tilDel_r & = \left(1-\lambda r^2\right) \left(r^2+\tila^2\right) - 2Mr \left(r^2+\tila^2\right)^{2-N} \,,\\[0.5cm]
	\tilSig & = r^2 + \tila^2\tilmu_0^2 \,.
\end{split}
\end{equation}

We see that the secondary directions enter the full metric only through the terms ${\bg_{_{\textrm{Eucl}}}}$ and $\tilPhi$.

Remembering that the coordinates are restricted by the constraint \eqref{eq:D=2N-sphere}, we conclude that the secondary block essentially obtains the geometry of a \linebreak($2N{-}3$)-dimen\-sional sphere. Indeed, it can be viewed as the sphere given by the constraint \eqref{eq:D=2N-sphere} embedded in a ($2N{-}2$)-dimensional flat space described by the metric \eqref{eq:D=2N-metric-flat}. Similarly to the general case, this sphere is coupled only to the primary coordinate $\tilmu_1$.

Besides this metric piece, the secondary directions enter the full metric also in time-related terms through the K\"{a}hler potential~$\tilPhi$. It is related to the common rotation of the secondary directions.

\subsection{Myers--Perry black hole in $\boldsymbol{D=6}$}

In this section, we simplify the situation even more and study the equal-spin limit of a rotating black hole in vacuum --- also called a Myers--Perry black hole --- in six dimensions. This was also discussed by Ortaggio \cite{Ortaggio2017}. Such a simplification proves useful as we are able to find additional Killing vectors that emerge after the limit, thus providing evidence of an enhanced symmetry structure of the resulting spacetime. We expect the symmetry group to become enlarged also in the general limit case.

We focus on a vacuum case, i.e. we set $\lambda=0$. Moreover, there are only two equal-spin blocks $\alpha\in\{1,\tilN\}$, and since in six dimensions a black hole described by the Lorentzian metric has only two rotational parameters, the first block has $\upind{1}N=1$ parameter subject to the limit. Therefore, the limiting procedure in this case is characterized by $a_{1,1}\rightarrow a_{1,0}$. Since this is a special case of the limit discussed in the previous subsection, we shall use the parametrization \eqref{eq:D=2N-parametrization} with the conventions \eqref{eq:D=2N-rename-a} and \eqref{eq:D=2N-rename-phi}
\begin{equation}
\label{eq:6D-parametrization}
\begin{array}{lll}
	a_{1,0} \equiv \tila \,, & x_{1,0} \equiv \tilx \,, & \phi_{1,0} \equiv \tilphi \,, \\[0.1cm]
	a_{1,1} \equiv \tila + a\varepsilon \,, \hphantom{a} & x_{1,1} \equiv \tila + x\varepsilon \,, \hphantom{a} & \phi_{1,1} \equiv \phi \,, \\[0.1cm]
	a_{\tilN,0} \equiv \tila_{\tilN} \,, & x_{\tilN,0} \equiv \tilx_{\tilN} \,, & \phi_{\tilN,0} \equiv \tilphi_{\tilN} \,,
\end{array}
\end{equation}
where we have also dropped the index $\rho=1$. Moreover, $\tilx_{\tilN}$ and $\tilphi_{\tilN}$ are Wick-rotated as in \eqref{eq:Wick-limit}, where we must perform the limit ${\lambda\to0}$ assuming the gauge condition \eqref{eq:gauge-limit} for $\tila_{\tilN}$.

\subsubsection*{Metric}

Employing these parametrizations, the metric \eqref{eq:metric-limit} becomes\\[-4ex]
\begin{equation}
\begin{split}
\label{eq:6D-metric-x}
	&\bg \approx  -\bd t^2 + \frac{2Mr}{\Sigma} \left[ \bd t - \frac{\tilx^2{-}\tila^2}{\tila} \left( \frac{x}{a}\, \bd\tilphi - \frac{x{-}a}{a}\, \bd\phi \right) \right]^2 \\[0.1cm]
	&\quad + \frac{\Sigma}{\Delta_r}\bd r^2 - \frac{r^2+\tilx^2}{\tilx^2-\tila^2}\bd\tilx^2 \\[0.2cm]
	&\quad + \frac{\left(r^2 {+} \tila^2\right)\left(\tilx^2{-}\tila^2\right)}{\tila^2}
	\left[ \frac{\bd x^2}{4x \left(x{-}a\right)} - \frac{x}{a} \bd\tilphi^2 + \frac{x{-}a}{a} \bd\phi^2 \right] \,,
\end{split}\raisetag{12ex}
\end{equation}
where
\begin{equation}
\begin{split}
	\Delta_r \approx & \left(r^2+\tila^2\right)\tilDel_r = \left(r^2+\tila^2\right)^2 - 2Mr \,, \\[0.2cm]
	\Sigma \approx & \left(r^2+\tila^2\right)\tilSig = \left(r^2+\tila^2\right) \left(r^2+\tilx^2\right) \,,
\end{split}
\end{equation}
and $\tilDel_r$ with $\tilSig$ are as in \eqref{eq:bh-func-mu-limit}, which confirms the results of \cite{Ortaggio2017}. This is the metric expressed in the generalized Boyer--Lindquist coordinates \{$t$, $r$, $\tilx$, $\tilphi$, $x$, $\phi$\}, which are suitable for performing the limiting procedure. However, considering the physical interpretation of the resulting spacetime, another set of coordinates proves more useful.

\subsubsection*{Spherical-like coordinates}

Instead of \{$\tilx$, $x$\} we introduce new angular coordinates \{$\vartheta$, $\chi$\}, which are better suited for analysing physical properties of the resulting black hole spacetime. Namely, let us define
\begin{equation}
\label{eq:6D-xx-thetachi}
	\tilx  = \tila\cos\vartheta \,,\qquad
	x  = a\cos^2\chi \,.
\end{equation}
Moreover, let us rename the angular coordinates \{$\tilphi$, $\phi$\} as follows
\begin{equation}
	\tilphi  \equiv \varphi_1 \,,\qquad
	\phi  \equiv \varphi_2 \,.
\end{equation}
The metric \eqref{eq:6D-metric-x} then becomes
\begin{equation}
\begin{split}
\label{eq:6D-metric-thetachi}
	\bg \approx & -\bd t^2 \\[0.2cm]
	& + \frac{2Mr}{\Sigma} \left[ \bd t + \tila\sin^2\vartheta \left( \cos^2\chi \bd\varphi_1 + \sin^2\chi \bd\varphi_2 \right) \right]^2 \\[0.2cm]
	& + \frac{\Sigma}{\Delta_r}\bd r^2 + \left(r^2+\tila^2\cos^2\vartheta\right) \bd\vartheta^2 \\[0.2cm]
	& + \left(r^2+\tila^2\right)\sin^2\vartheta\,\bd\mathbf{S}_3 \,,
\end{split}\raisetag{11ex}
\end{equation}
where $\bd\mathbf{S}_3$ denotes the metric of a 3-sphere, which can be expressed as\footnote{The coordinates in which the metric of a 3-sphere has this particular form are called the \emph{Hopf coordinates}.}
\begin{equation}
\label{eq:3-sphere}
	\bd\mathbf{S}_3 = \bd\chi^2 + \cos^2\chi\, \bd\varphi_1^2 + \sin^2\chi\, \bd\varphi_2^2 \,.
\end{equation}
The metric functions are
\begin{equation}
\begin{split}
	\Delta_r \approx & \left(r^2+\tila^2\right)^2 - 2Mr \,, \\[0.2cm]
	\Sigma \approx & \left(r^2+\tila^2\right) \left(r^2+\tila^2\cos^2\vartheta\right) \,.
\end{split}
\end{equation}
Notice that the metric \eqref{eq:6D-metric-thetachi} no longer contains the parameter $a$ from the secondary block. This parameter controls how fast $a_{1,1}$ approaches $a_{1,0}$. In case the limit is applied to a single rotational parameter, however, such a scale is irrelevant.

The inverse metric can be written as
\begin{equation}
\begin{split}
	\bg^{-1} & \approx
	- \left(\framevec{t}\right)^2 \\[0.2cm]
	& - \frac{2Mr}{\Sigma}\frac{\left(r^2+\tila^2\right)^2}{\Delta_r} \left[\framevec{t} - \frac{\tila}{r^2+\tila^2} \left(\framevec{\varphi_1} + \framevec{\varphi_2}\right)\right]^2 \\[0.2cm]
	& + \frac{\Delta_r}{\Sigma}\left(\framevec{r}\right)^2 + \frac{1}{r^2+\tila^2\cos^2\vartheta} \left(\framevec{\vartheta}\right)^2 \\[0.2cm]
	& + \frac{1}{\left(r^2+\tila^2\right)\sin^2\vartheta} \bd\mathbf{S}_3^{-1} \,,
\end{split}
\end{equation}
with the inverse metric of a 3-sphere given simply by
\begin{equation}
	\bd\mathbf{S}_3^{-1} = \left(\framevec{\chi}\right)^2 + \frac{1}{\cos^2\chi} \left(\framevec{\varphi_1}\right)^2 + \frac{1}{\sin^2\chi} \left(\framevec{\varphi_2}\right)^2 \,.
\end{equation}

\subsubsection*{Killing vectors}

Before the limit, the Myers--Perry black hole in six dimensions has three explicit symmetries associated with the following Killing vectors
\begin{equation}
\label{eq:6D-killvect-orig}
\begin{split}
	\bxi & = \framevec{t} \,,\\[0.2cm]
	\bs_+ & = \frac{1}{2} \left(\framevec{\varphi_1}+\framevec{\varphi_2}\right)
    =\frac{1}{2\lambda}\left(\frac{\bs_{(1)}}{a_1}+\frac{\bs_{(2)}}{a_2}\right) \,,\\[0.2cm]
	\bs_- & = \frac{1}{2} \left(\framevec{\varphi_1}-\framevec{\varphi_2}\right)
    =\frac{1}{2\lambda}\left(\frac{\bs_{(1)}}{a_1}-\frac{\bs_{(2)}}{a_2}\right) \,,
\end{split}
\end{equation}
where $\bxi$ is the special Killing vector \eqref{xidef}. Note that $\bs_+$ and $\bs_-$ are simply linear combinations of the vectors $\bs_{(1)}$ and $\bs_{(2)}$ defined in \eqref{eq:s-def}\footnote{Let us note that the factor $1/\lambda$ in the expressions for $\bs_\pm$ in terms of $\bs_{(1,2)}$ does not present a problem in the limit $\lambda\to0$ since it is compensated by $\lambda$ in the definition \eqref{eq:s-def}.}.  However, they form nice commutation relations with new Killing vectors, which we will identify next.

The most important result of this section is that the metric after the limit has additional Killing vectors. Namely, we found two new vectors
\begin{equation}
\begin{split}
	\boldsymbol{u} & = \frac{1}{2} \left[ \cos(\varphi_2-\varphi_1) \framevec{\chi} \right. \\[0.1cm]
	& \hspace{1.5cm} \left. - \sin(\varphi_2-\varphi_1) \left( \tan\chi \framevec{\varphi_1} + \cot\chi \framevec{\varphi_2} \right) \right] \,,\\[0.2cm]
	\boldsymbol{v} & = \frac{1}{2} \left[ \sin(\varphi_2-\varphi_1) \framevec{\chi} \right. \\[0.1cm]
	& \hspace{1.5cm} \left. + \cos(\varphi_2-\varphi_1) \left( \tan\chi \framevec{\varphi_1} + \cot\chi \framevec{\varphi_2} \right) \right] \,.
\end{split}\raisetag{14ex}
\end{equation}
These vectors are independent of the original Killing vectors \eqref{eq:6D-killvect-orig} and they Lie-preserve the full spacetime metric \eqref{eq:6D-metric-thetachi}. Furthermore, it can be shown that they are Killing vectors of a 3-sphere represented by the metric \eqref{eq:3-sphere}.

Let us now discuss how the symmetry group changes after applying the limiting procedure. Before the limit, the spacetime symmetries
form a $\realn\times U(1)\times U(1)$ group as there exist three commuting Killing vectors $\bxi$, $\bs_+$ and $\bs_-$. However, the algebraic structure emerging after the limit indicates that the symmetry of the resulting spacetime is indeed further enhanced. In fact, the vectors $\bs_-$, $\boldsymbol{u}$ and $\boldsymbol{v}$ generate the algebra of an $SO(3)$ group and the vectors $\bxi$ and $\bs_+$ commute with all the other vectors as can be seen from their Lie brackets\footnote{Commutation relations with $\bxi$ are trivial since all the other Killing vectors are time-independent.}
\begin{equation}
\begin{split}
	\left[\boldsymbol{s}_-,\boldsymbol{u}\right] & = \boldsymbol{v} \,,\qquad
	\left[\boldsymbol{u},\boldsymbol{v}\right] = \boldsymbol{s}_- \,,\qquad
	\left[\boldsymbol{v},\boldsymbol{s}_-\right] = \boldsymbol{u} \,, \\[0.2cm]
	\left[\boldsymbol{s}_+,\boldsymbol{s}_-\right] & = 0 \,,\quad\;\;
	\left[\boldsymbol{s}_+,\boldsymbol{u}\right] = 0 \,,\qquad\;\;\,
	\left[\boldsymbol{s}_+,\boldsymbol{v}\right] = 0 \,.
\end{split}
\end{equation}
Therefore, the symmetry group of the spacetime decouples and is enhanced from the original $\realn\times U(1)\times U(1)$ to $\realn\times U(1)\times SO(3)$ after the limit.

\subsubsection*{Killing tensors}

Another proof of an enhanced symmetry structure after performing the equal-spin limit can be found when studying the limit form of Killing tensors. In six dimensions, the Myers--Perry black hole after the limit has three Killing tensors defined in \eqref{eq:K-limit} and \eqref{eq:k-secondary}. In the spherical-like coordinates, they obtain the form\footnote{Here, $\ts{\alpha}\vee\ts{\beta}=\ts{\alpha}\ts{\beta}+\ts{\beta}\ts{\alpha}$ is a normalized symmetric tensor product analogous to the antisymmetric wedge operation.}
\begin{equation}
\begin{split}
	&\tilK_{(0)} \approx \bg^{-1} \,,\\[1.5ex]
	&\tilK_{(1)}  = -\left(\tila\framevec{t}\right)^{\!2}
    + \tila\,\framevec{t}\vee \left(\framevec{\varphi_1} + \framevec{\varphi_2}\right) \\[1ex]
	& \;\; - \frac{2Mr}{\Sigma} \frac{\left[\bigl(r^2{+}\tila^2\bigr) \tila\cos\vartheta\right]^2}{\Delta_r}\!
    \left[\framevec{t} {-} \frac{\tila}{r^2{+}\tila^2} \Bigl(\framevec{\varphi_1} {+} \framevec{\varphi_2}\Bigr)\right]^2 \\[1ex]
	& \;\; + \frac{\Delta_r}{\Sigma}\left(\tila\cos\vartheta\framevec{r}\right)^{\!2}
    - \frac{1}{r^2+\tila^2\cos^2\vartheta} \left(r\framevec{\vartheta}\right)^{\!2} \\[1ex]
	& \;\; - \frac{r^2+\tila^2\sin^2\vartheta}{\left(r^2+\tila^2\right)\sin^2\vartheta} \bd\mathbf{S}_3^{-1} \,,\\[1.5ex]
	&\upind{1}\bk_{(0)} = \upind{1}\bg^{-1} = \frac{1}{2} \left[ \Bigl(\framevec{\chi}\Bigr)^2
    + \Bigl(\tan\chi\framevec{\varphi_1} {-} \cot\chi\framevec{\varphi_2}\Bigr)^2 \right] \,,
\end{split}\raisetag{23ex}
\end{equation}
where $\upind{1}\bg^{-1}$ is the K\"{a}hler part of the full spacetime metric $\bg$. It turns out that $\upind{1}\bk_{(0)}$ decouples into a sum of direct products of Killing vectors in the following manner
\begin{equation}
	\upind{1}\bk_{(0)} = 2 \left( \boldsymbol{u}\boldsymbol{u} + \boldsymbol{v}\boldsymbol{v} + \bs_-\bs_- - \bs_+\bs_+ \right) \,,
\end{equation}
thus the Killing tensor $\upind{1}\bk_{(0)}$ becomes reducible. Therefore, the corresponding hidden symmetry splits into a combination of explicit symmetries characterized by the Killing vectors $\bs_+$, $\bs_-$, $\boldsymbol{u}$ and $\boldsymbol{v}$.


\section{Summary}
\label{sc:summary}

This work is devoted to a limit of the Kerr--NUT--(A)dS spacetime, in which an arbitrary number of its rotational parameters coincides. The importance of the Kerr--NUT--(A)dS spacetime lies in the fact that it is the most general solution to the vacuum Einstein equations in higher dimensions with the cosmological constant that also possesses the principal tensor. Therefore, it has a rich symmetry structure demonstrated by the existence of the Killing tower of Killing vectors and Killing tensors. The limiting spacetime inherits this symmetry structure. The symmetry is even enhanced since some of the hidden symmetries represented by Killing tensors factorize, which leads to a higher number of explicit symmetries represented by Killing vectors.

Section \ref{sc:metric-limit} describes the equal-spin limit of the Kerr--(A)dS metric. Although the limiting procedure is not trivial since some of the parameters and the coordinate ranges become degenerate, we managed to find a suitable parametrization of the limit, in which the metric remains regular. We defined primary and secondary \emph{coordinate} directions, which refer to the coordinates (and the parameters) that do not change after the limit and to those subject to the limit, respectively. We then applied the limiting procedure to the orthogonal frame \{$\beps^{\mu}$, $\hat{\beps}^{\mu}$\} and found out that after the limit this orthogonal frame separates into two independent sets \{$\tilbeps^{\alpha}$, $\htilbeps^{\alpha}$\} and \{$\upal\beps^{\rho}$, \nolinebreak $\upal\hatbeps^{\rho}$\}, which we referred to as primary and secondary \emph{frame} directions, respectively.

This structure corresponds to the expected form of the generalized Kerr--NUT--(A)dS spacetimes \cite{Houri2008}, where spacetimes with the principal tensor that has constant eigenvalues have been discussed. In \cite{Houri2008}, the most general metric allowing such a principal tensor has been identified. It was shown that certain parts of the metric have formal properties of the K\"ahler metrics, which we confirmed in our limit. We thus kept the terminology of~\cite{Houri2008}.

However, the orthogonal separation of the metric is only valid on the level of tangent spaces and is not integrable. Thus the resulting geometry cannot be understood as a product of independent manifolds and one cannot talk about true K\"ahler submanifolds. The resulting geometry is a certain kind of multi-warped product.

Applying the limit we discovered that the metric ${\bg}$ splits into two parts --- the primary Kerr--NUT--(A)dS part $\tilbg$ and the secondary K\"{a}hler metrics $\upal\bg$. The Kerr--NUT--(A)dS part has the form analogous to the Kerr--NUT--(A)dS metric of the entire spacetime but only in the primary directions, while the K\"{a}hler metrics describe special parts of the geometry that emerged in the limiting procedure. However, both pieces are interlaced together because of the fine inner structure of the primary and secondary orthogonal frames. Thus, we repeat once more, that the secondary metrics $\upal\bg$ can be called K\"{a}hler metrics only formally since they do not live on independent submanifolds.

As a particular case we studied the limit of a black hole geometry with the explicit Lorentzian signature. In this case, the interlacing of the primary Kerr--NUT--(A)dS block and the secondary K\"{a}hler blocks can be demonstrated in an explicit form. It turns out that after the limit the secondary blocks simplify to spatial spheres, and each sphere is coupled to the coordinate in the corresponding primary direction. Additionally, the secondary blocks are coupled to the primary metric in the rotational term linear in ${\bd t}$ through the K\"ahler potentials.

In Section \ref{sc:symmetries-limit}, we discussed the equal-spin limit of the Killing tower. We were able to obtain the same number of Killing vectors and Killing tensors as before the limit, thus having reconstructed the original explicit and hidden symmetries.

Therefore, the metric we obtained represents a particularly interesting subcase of general spacetimes discussed in \cite{Houri2008}. We showed that it retains the full tower of explicit and hidden symmetries, thus inheriting the complete integrability of the geodesic particle motion from the original geometry --- unlike the generalized Kerr--NUT--(A)dS metrics, which do not necessarily admit the full Killing tower and they possess much less symmetries.

In general, we expect that the limiting metric has even enhanced symmetry structure, i.e. that it possesses more explicit symmetries, making some of the hidden symmetries reducible.

In Section \ref{sc:examples}, we presented two examples of the general results. We studied the metric of a black hole with all its rotational parameters set equal in a general (even) dimension and, in particular, in six dimensions. In the six-dimensional case, we found an enhanced symmetry structure after the limit. Namely, we discovered two additional Killing vectors that emerge after performing the limiting procedure. These new vectors are independent of the original Killing vectors and they Lie-preserve the full spacetime metric. Moreover, combined with the original Killing vectors they generate the algebra of an $SO(3)$ group. Therefore, the symmetry group of the spacetime is enhanced from the original $\realn\times U(1)\times U(1)$ to $\realn\times U(1)\times SO(3)$ after the limit. It also turns out that one of the Killing tensors becomes reducible as it decouples into a sum of products of Killing vectors, therefore, the associated hidden symmetry splits into a combination of explicit symmetries.

Let us conclude with several open problems concerning the equal-spin limit of the Kerr--NUT--(A)dS spacetime.

We expect the symmetry group to become enlarged in the general limit case. Namely, we expect that after the limit the symmetries of spherically symmetric spatial parts emerge on the level of the full spacetime metric. This should be possible to prove using a higher-dimensional generalization of the spherical-like coordinates \eqref{eq:6D-xx-thetachi}. While the task is rather simple when the secondary block contains a single coordinate, with more coordinates the complexity of the expressions grows considerably.

In our work, we mostly assumed vanishing NUT charges for simplicity. Some of the results could be generalized to the case of non-vanishing NUT charges, but it would require much more careful fine-tuning of the metric parameters and the coordinate ranges and one would need to investigate the relation of the resulting symmetry structure with the NUT-related singular structure of the axes.

\section*{Acknowledgements}

The work was supported by the Czech Science Foundation Grant 20-05421S. E.~P.\ thanks also the Grant Agency of Charles University for the support under the project GAUK 514218.


\appendix


\section{Metric functions}
\label{app:metric-funcs-def}

In Section \ref{sc:kerr-nut-ads}, we have introduced several auxiliary functions related to symmetric polynomials that are then often used in the metric and other quantities before the limit, and with some modifications also after the limit (see Appendix \ref{app:metric-funcs-limit}). These functions are polynomials either in the coordinates $x_{\mu}$ (denoted by $J$, $A$, $U$) or in the parameters~$a_{\mu}$ (denoted by $\calJ$, $\calA$, $\calU$). They are defined as%
\footnote{Let us remind that indices in sums and products go over the ``default'' ranges unless indicated otherwise explicitly, i.e.
\begin{equation*}
	\sum_{\mu} = \sum_{\mu=1}^N \,,\qquad \sum_k = \sum_{k=0}^{N-1} \,.
\end{equation*}
}
\begin{equation}
\label{eq:app-J}
\begin{split}
	J(a^2) & = \prod_{\nu}\left(x_{\nu}^2-a^2\right) = \sum_{k=0}^{N} A^{(k)}(-a^2)^{N-k} \,, \\[0.2cm]
	\calJ(x^2) & = \prod_{\nu}\left(a_{\nu}^2-x^2\right) = \sum_{k=0}^{N} \calA^{(k)}(-x^2)^{N-k} \,.
\end{split}
\end{equation}
Considering these definitions, it follows that
\begin{equation}
\label{eq:app-A}
	A^{(k)} = \mkern-18mu \sum_{\overset{\scriptstyle{\nu_1,\dots,\nu_k}}{\scriptstyle{\nu_1<\dots<\nu_k}}} \mkern-18mu x_{\nu_1}^2\dots x_{\nu_k}^2 \,,\quad
	\calA^{(k)} = \mkern-18mu \sum_{\overset{\scriptstyle{\nu_1,\dots,\nu_k}}{\scriptstyle{\nu_1<\dots<\nu_k}}} \mkern-18mu a_{\nu_1}^2\dots a_{\nu_k}^2 \,.
\end{equation}
Similarly, we define functions with the $\mu$-th variable omitted as
\begin{equation}
\label{eq:app-Jmu}
\begin{split}
	J_{\mu}(a^2) & = \prod_{\overset{\scriptstyle{\nu}}{\scriptstyle{\nu\neq\mu}}} \left(x_{\nu}^2-a^2\right) = \sum_k A_{\mu}^{(k)}(-a^2)^{N-1-k} \,, \\[0.2cm]
	\calJ_{\mu}(x^2) & = \prod_{\overset{\scriptstyle{\nu}}{\scriptstyle{\nu\neq\mu}}} \left(a_{\nu}^2-x^2\right) = \sum_k \calA_{\mu}^{(k)}(-x^2)^{N-1-k} \,,
\end{split}
\end{equation}\\[-3ex]
where
\begin{equation}
\label{eq:app-Amu}
	A_{\mu}^{(k)} = \mkern-18mu \sum_{\overset{\overset{\scriptstyle{\nu_1,\dots,\nu_k}}{\scriptstyle{\nu_1<\dots<\nu_k}}}{\scriptstyle{\nu_i\neq\mu}}} \mkern-18mu x_{\nu_1}^2\dots x_{\nu_k}^2 \,,\quad
	\calA_{\mu}^{(k)} = \mkern-18mu \sum_{\overset{\overset{\scriptstyle{\nu_1,\dots,\nu_k}}{\scriptstyle{\nu_1<\dots<\nu_k}}}{\scriptstyle{\nu_i\neq\mu}}} \mkern-18mu a_{\nu_1}^2\dots a_{\nu_k}^2 \,.
\end{equation}
We set
\begin{equation}
\label{eq:app-A0}
	A^{(0)} = A_{\mu}^{(0)} = 1 \,,\qquad \calA^{(0)} = \calA_{\mu}^{(0)} = 1
\end{equation}
and we also assume that the functions $A_{\mu}^{(k)}$ vanish if the index~$k$ ``overflows'', e.g., $A_{\mu}^{(N)}=0$; the same applies to the functions $\calA_{\mu}^{(k)}$. Finally, we define
\begin{equation}
\label{eq:app-U}
\begin{split}
	U_{\mu} & = J_{\mu}(x_{\mu}^2) = \prod_{\overset{\scriptstyle{\nu}}{\scriptstyle{\nu\neq\mu}}} \left(x_{\nu}^2-x_{\mu}^2\right) \,, \\[0.2cm]
	\calU_{\mu} & = \calJ_{\mu}(a_{\mu}^2) = \prod_{\overset{\scriptstyle{\nu}}{\scriptstyle{\nu\neq\mu}}}\left(a_{\nu}^2-a_{\mu}^2\right) \,.
\end{split}
\end{equation}

These functions satisfy
\begin{equation}
\begin{array}{rrr}
	J(x_{\mu}^2) = 0 \,,\quad\quad & \calJ(a_{\mu}^2) = 0 \,, \\[0.2cm]
	J_{\mu}(x_{\nu}^2) = 0 \,,\quad\quad & \calJ_{\mu}(a_{\nu}^2) = 0 \,, & \quad \textrm{if } \mu\neq\nu \,.
\end{array}
\end{equation}
If $A_{\mu}^{(k)}$ is understood as an ${N{\times}N}$ matrix, it is possible to write down its inverse
\begin{align}
	& \sum_k A_{\mu}^{(k)} \frac{(-x_{\nu}^2)^{N-1-k}}{U_{\nu}} = \delta_{\mu\nu} \,, \label{eq:id-A-sum-k} \\[0.2cm]
	& \sum_{\mu} A_{\mu}^{(k)} \frac{(-x_{\mu}^2)^{N-1-l}}{U_{\mu}} = \delta_{kl} \,. \label{eq:id-A-sum-mu}
\end{align}
They are also related to the polynomials $A^{(k)}$ as
\begin{equation}
	\sum_{\mu} \frac{A_{\mu}^{(k)}}{x_{\mu}^2 U_{\mu}} = \frac{A^{(k)}}{A^{(N)}} \,. \label{eq:id-AxU}
\end{equation}
The functions $\calA^{(k)}$ and $\calA_{\mu}^{(k)}$ satisfy analogous identities with $x_{\mu}$ and $U_{\mu}$ replaced by $a_{\mu}$ and $\calU_{\mu}$, respectively.

Finally, the following orthogonality relations are satisfied
\begin{align}
	\sum_{\kappa} \frac{J_{\mu}(a_{\kappa}^2)}{\calU_{\kappa}} \frac{\calJ_{\kappa}(x_{\nu}^2)}{U_{\nu}} & = \delta_{\mu\nu} \,, \label{eq:og-JcalUcalJU} \\[0cm]
	\sum_{\kappa} \frac{J_{\mu}(a_{\kappa}^2)J_{\nu}(a_{\kappa}^2)}{J(a_{\kappa}^2)\,\calU_{\kappa}} & = -\frac{U_{\mu}}{\calJ(x_{\mu}^2)}\,\delta_{\mu\nu} \,, \label{eq:og-JJJcalU} \\[0cm]
	\sum_{\kappa} J_{\kappa}(a_{\mu}^2)J_{\kappa}(a_{\nu}^2) \, \frac{\calJ(x_{\kappa}^2)}{U_{\kappa}} & = -J(a_{\mu}^2)\,\calU_{\mu}\,\delta_{\mu\nu} \,. \label{eq:og-JJcalJU}
\end{align}\\[-2ex]


\section{Equal-spin limit of metric functions}
\label{app:metric-funcs-limit}

When employing the limiting procedure introduced in Section \ref{sc:metric-limit}, first we rewrite the metric functions using the double indexing described in subsection \ref{sc:double-indexing}, and then apply the parametrization \eqref{eq:parametrization}. The functions $J$ and $U$ along with their counterparts $\calJ$ and $\calU$ can be written as%
\footnote{Let us remind that indices introduced in Section \ref{sc:double-indexing} go over the ``default'' ranges unless indicated otherwise explicitly in the sum or product, i.e.,
\begin{equation*}
	\prod_{\alpha} = \prod_{\alpha=1}^{\tilN} \,,\qquad \upal\!\prod_{\rho} = \prod_{\rho=1}^{\upal N} \,.
\end{equation*}
}
\begin{widetext}
\begingroup
\allowdisplaybreaks
\begin{align}
	\calJ(x_{\alpha,0}^2) & = \prod_{\gamma} \left(a_{\gamma,0}^2-x_{\alpha,0}^2\right) \;\cdot\; \prod_{\gamma}\; \upind{\gamma}\!\prod_{\tau} \left(a_{\gamma,\tau}^2-x_{\alpha,0}^2\right) \,,\\[0.0mm]
	\calJ(x_{\alpha,\rho}^2) & = \prod_{\gamma} \left(a_{\gamma,0}^2-x_{\alpha,\rho}^2\right) \;\cdot\; \prod_{\gamma}\; \upind{\gamma}\!\prod_{\tau} \left(a_{\gamma,\tau}^2-x_{\alpha,\rho}^2\right) \,,\\[0.0mm]
	J_{\alpha,0}(a_{\beta,0}^2) & = \prod_{\overset{\scriptstyle{\gamma}}{\scriptstyle{\gamma\neq\alpha}}} \left(x_{\gamma,0}^2-a_{\beta,0}^2\right) \;\cdot\; \prod_{\gamma}\; \upind{\gamma}\!\prod_{\tau} \left(x_{\gamma,\tau}^2-a_{\beta,0}^2\right) \,,\\[0.0mm]
	J_{\alpha,0}(a_{\beta,\sigma}^2) & = \prod_{\overset{\scriptstyle{\gamma}}{\scriptstyle{\gamma\neq\alpha}}} \left(x_{\gamma,0}^2-a_{\beta,\sigma}^2\right) \;\cdot\; \prod_{\gamma}\; \upind{\gamma}\!\prod_{\tau} \left(x_{\gamma,\tau}^2-a_{\beta,\sigma}^2\right) \,,\\[0.0mm]
	J_{\alpha,\rho}(a_{\beta,0}^2) & = \prod_{\gamma} \left(x_{\gamma,0}^2-a_{\beta,0}^2\right) \;\cdot\; \prod_{\overset{\scriptstyle{\gamma}}{\scriptstyle{\gamma\neq\alpha}}} \; \upind{\gamma}\!\prod_{\tau} \left(x_{\gamma,\tau}^2-a_{\beta,0}^2\right) \;\cdot\; \upind{\alpha}\! \prod_{\overset{\scriptstyle{\tau}}{\scriptstyle{\tau\neq\rho}}} \left(x_{\alpha,\tau}^2-a_{\beta,0}^2\right) \,, \\[0.0mm]
	J_{\alpha,\rho}(a_{\beta,\sigma}^2) & = \prod_{\gamma} \left(x_{\gamma,0}^2-a_{\beta,\sigma}^2\right) \;\cdot\; \prod_{\overset{\scriptstyle{\gamma}}{\scriptstyle{\gamma\neq\alpha}}} \; \upind{\gamma}\!\prod_{\tau} \left(x_{\gamma,\tau}^2-a_{\beta,\sigma}^2\right) \;\cdot\; \upind{\alpha} \! \prod_{\overset{\scriptstyle{\tau}}{\scriptstyle{\tau\neq\rho}}} \left(x_{\alpha,\tau}^2-a_{\beta,\sigma}^2\right) \,, \\[0.0mm]
	\calJ_{\beta,0}(x_{\alpha,0}^2) & = \prod_{\overset{\scriptstyle{\gamma}}{\scriptstyle{\gamma\neq\beta}}} \left(a_{\gamma,0}^2-x_{\alpha,0}^2\right) \;\cdot\; \prod_{\gamma}\; \upind{\gamma}\!\prod_{\tau} \left(a_{\gamma,\tau}^2-x_{\alpha,0}^2\right) \,,\\[0.0mm]
	\calJ_{\beta,\sigma}(x_{\alpha,0}^2) & = \prod_{\gamma} \left(a_{\gamma,0}^2-x_{\alpha,0}^2\right) \;\cdot\; \prod_{\overset{\scriptstyle{\gamma}}{\scriptstyle{\gamma\neq\beta}}} \; \upind{\gamma}\!\prod_{\tau} \left(a_{\gamma,\tau}^2-x_{\alpha,0}^2\right) \;\cdot\; \upind{\beta} \! \prod_{\overset{\scriptstyle{\tau}}{\scriptstyle{\tau\neq\sigma}}} \left(a_{\beta,\tau}^2-x_{\alpha,0}^2\right) \,, \\[0.0mm]
	\calJ_{\beta,0}(x_{\alpha,\rho}^2) & = \prod_{\overset{\scriptstyle{\gamma}}{\scriptstyle{\gamma\neq\beta}}} \left(a_{\gamma,0}^2-x_{\alpha,\rho}^2\right) \;\cdot\; \prod_{\gamma}\upind{\gamma}\prod_{\tau} \left(a_{\gamma,\tau}^2-x_{\alpha,\rho}^2\right) \,,\\[0.0mm]
	\calJ_{\beta,\sigma}(x_{\alpha,\rho}^2) & = \prod_{\gamma} \left(a_{\gamma,0}^2-x_{\alpha,\rho}^2\right) \;\cdot\; \prod_{\overset{\scriptstyle{\gamma}}{\scriptstyle{\gamma\neq\beta}}} \; \upind{\gamma}\!\prod_{\tau} \left(a_{\gamma,\tau}^2-x_{\alpha,\rho}^2\right) \;\cdot\; \upind{\beta}\!\prod_{\overset{\scriptstyle{\tau}}{\scriptstyle{\tau\neq\sigma}}} \left(a_{\beta,\tau}^2-x_{\alpha,\rho}^2\right) \,, \\[0.0mm]
	U_{\alpha,0} & = \prod_{\overset{\scriptstyle{\gamma}}{\scriptstyle{\gamma\neq\alpha}}} \left(x_{\gamma,0}^2-x_{\alpha,0}^2\right) \;\cdot\; \prod_{\gamma}\; \upind{\gamma}\!\prod_{\tau} \left(x_{\gamma,\tau}^2-x_{\alpha,0}^2\right) \,,\\[0.0mm]
	U_{\alpha,\rho} & = \prod_{\gamma} \left(x_{\gamma,0}^2-x_{\alpha,\rho}^2\right) \;\cdot\; \prod_{\overset{\scriptstyle{\gamma}}{\scriptstyle{\gamma\neq\alpha}}} \; \upind{\gamma}\!\prod_{\tau} \left(x_{\gamma,\tau}^2-x_{\alpha,\rho}^2\right) \;\cdot\;\upind{\alpha}\!\prod_{\overset{\scriptstyle{\tau}}{\scriptstyle{\tau\neq\rho}}} \left(x_{\alpha,\tau}^2-x_{\alpha,\rho}^2\right) \,,\\[0.0mm]
	\calU_{\alpha,0} & = \prod_{\overset{\scriptstyle{\gamma}}{\scriptstyle{\gamma\neq\alpha}}} \left(a_{\gamma,0}^2-a_{\alpha,0}^2\right) \;\cdot\; \prod_{\gamma}\; \upind{\gamma}\!\prod_{\tau} \left(a_{\gamma,\tau}^2-a_{\alpha,0}^2\right) \,,\\[0.0mm]
	\calU_{\alpha,\rho} & = \prod_{\gamma} \left(a_{\gamma,0}^2-a_{\alpha,\rho}^2\right) \;\cdot\; \prod_{\overset{\scriptstyle{\gamma}}{\scriptstyle{\gamma\neq\alpha}}} \; \upind{\gamma}\!\prod_{\tau} \left(a_{\gamma,\tau}^2-a_{\alpha,\rho}^2\right) \;\cdot\; \upind{\alpha}\!\prod_{\overset{\scriptstyle{\tau}}{\scriptstyle{\tau\neq\rho}}} \left(a_{\alpha,\tau}^2-a_{\alpha,\rho}^2\right) \,.
\end{align}\\[-3ex]
After the limit, these functions become
\begin{align}
	\calJ(x_{\alpha,0}^2) & \approx \tilcalJ(\tilx_{\alpha}^2) \prod_{\gamma} \left(\tila_{\gamma}^2-\tilx_{\alpha}^2\right)^{\upga N} \,,\\[0mm]
	\calJ(x_{\alpha,\rho}^2) & \approx -\tilcalU_{\alpha} \upal x_{\rho}\upal\calJ(\upal x_{\rho}) \left(2\tila_{\alpha}\varepsilon\right)^{\upal N+1} \prod_{\overset{\scriptstyle{\gamma}}{\scriptstyle{\gamma\neq\alpha}}} \left(\tila_{\gamma}^2-\tila_{\alpha}^2\right)^{\upga N} \,,\\[0.0mm]
	J_{\alpha,0}(a_{\beta,0}^2) & \approx \tilJ_{\alpha}(\tila_{\beta}^2) \upbe J(0) \left(2\tila_{\beta}\varepsilon\right)^{\upbe N} \prod_{\overset{\scriptstyle{\gamma}}{\scriptstyle{\gamma\neq\beta}}} \left(\tila_{\gamma}^2-\tila_{\beta}^2\right)^{\upind{\gamma}N} \,,\\[0mm]
	J_{\alpha,0}(a_{\beta,\sigma}^2) & \approx \tilJ_{\alpha}(\tila_{\beta}^2) \upbe J(\upbe a_{\sigma}) \left(2\tila_{\beta}\varepsilon\right)^{\upbe N} \prod_{\overset{\scriptstyle{\gamma}}{\scriptstyle{\gamma\neq\beta}}} \left(\tila_{\gamma}^2-\tila_{\beta}^2\right)^{\upind{\gamma}N} \,,\\[-0.2mm]
	J_{\alpha,\rho}(a_{\beta,0}^2) & \approx \frac{\tilJ(\tila_{\beta}^2)}{\tila_{\alpha}^2-\tila_{\beta}^2} \upbe J(0) \left(2\tila_{\beta}\varepsilon\right)^{\upbe N} \prod_{\overset{\scriptstyle{\gamma}}{\scriptstyle{\gamma\neq\beta}}} \left(\tila_{\gamma}^2-\tila_{\beta}^2\right)^{\upind{\gamma}N} \,,\qquad \alpha\neq\beta \,, \\[-0.2mm]
	J_{\alpha,\rho}(a_{\beta,\sigma}^2) & \approx \frac{\tilJ(\tila_{\beta}^2)}{\tila_{\alpha}^2-\tila_{\beta}^2} \upbe J(\upbe a_{\sigma}) \left(2\tila_{\beta}\varepsilon\right)^{\upbe N} \prod_{\overset{\scriptstyle{\gamma}}{\scriptstyle{\gamma\neq\beta}}} \left(\tila_{\gamma}^2-\tila_{\beta}^2\right)^{\upind{\gamma}N} \,,\qquad \alpha\neq\beta \,, \\[0mm]
	J_{\alpha,\rho}(a_{\alpha,0}^2) & \approx \tilJ(\tila_{\alpha}^2) \upal J_{\rho}(0) \left(2\tila_{\alpha}\varepsilon\right)^{\upal N-1} \prod_{\overset{\scriptstyle{\gamma}}{\scriptstyle{\gamma\neq\alpha}}} \left(\tila_{\gamma}^2-\tila_{\alpha}^2\right)^{\upga N} \,, \\[0.0mm]
	J_{\alpha,\rho}(a_{\alpha,\sigma}^2) & \approx \tilJ(\tila_{\alpha}^2) \upal J_{\rho}(\upal a_{\sigma}) \left(2\tila_{\alpha}\varepsilon\right)^{\upal N-1} \prod_{\overset{\scriptstyle{\gamma}}{\scriptstyle{\gamma\neq\alpha}}} \left(\tila_{\gamma}^2-\tila_{\alpha}^2\right)^{\upga N} \,, \\[0.0mm]
	\calJ_{\beta,0}(x_{\alpha,0}^2) & \approx \tilcalJ_{\beta}(\tilx_{\alpha}^2) \prod_{\gamma} \left(\tila_{\gamma}^2-\tilx_{\alpha}^2\right)^{\upind{\gamma}N} \,,\\[0.0mm]
	\calJ_{\beta,\sigma}(x_{\alpha,0}^2) & \approx \tilcalJ_{\beta}(\tilx_{\alpha}^2) \prod_{\gamma} \left(\tila_{\gamma}^2-\tilx_{\alpha}^2\right)^{\upind{\gamma}N} \,, \\[0.0mm]
	\calJ_{\beta,0}(x_{\alpha,\rho}^2) & \approx \frac{\tilcalU_{\alpha}}{\tila_{\alpha}^2-\tila_{\beta}^2} \upal x_{\rho}\upal\calJ(\upal x_{\rho}) \left(2\tila_{\alpha}\varepsilon\right)^{\upal N+1} \prod_{\overset{\scriptstyle{\gamma}}{\scriptstyle{\gamma\neq\alpha}}} \left(\tila_{\gamma}^2-\tila_{\alpha}^2\right)^{\upind{\gamma} N} \,,\qquad \alpha\neq\beta \,, \\[0.0mm]
	\calJ_{\beta,\sigma}(x_{\alpha,\rho}^2) & \approx \frac{\tilcalU_{\alpha}}{\tila_{\alpha}^2-\tila_{\beta}^2} \upal x_{\rho}\upal\calJ(\upal x_{\rho}) \left(2\tila_{\alpha}\varepsilon\right)^{\upal N+1} \prod_{\overset{\scriptstyle{\gamma}}{\scriptstyle{\gamma\neq\alpha}}} \left(\tila_{\gamma}^2-\tila_{\alpha}^2\right)^{\upind{\gamma} N} \,,\qquad \alpha\neq\beta \,, \\[0.0mm]
	\calJ_{\alpha,0}(x_{\alpha,\rho}^2) & \approx \tilcalU_{\alpha} \upal\calJ(\upal x_{\rho}) \left(2\tila_{\alpha}\varepsilon\right)^{\upal N} \prod_{\overset{\scriptstyle{\gamma}}{\scriptstyle{\gamma\neq\alpha}}} \left(\tila_{\gamma}^2-\tila_{\alpha}^2\right)^{\upind{\gamma} N} \,, \\[0.0mm]
	\calJ_{\alpha,\sigma}(x_{\alpha,\rho}^2) & \approx -\tilcalU_{\alpha} \upal x_{\rho} \upal\calJ_{\sigma}(\upal x_{\rho}) \left(2\tila_{\alpha}\varepsilon\right)^{\upal N} \prod_{\overset{\scriptstyle{\gamma}}{\scriptstyle{\gamma\neq\alpha}}} \left(\tila_{\gamma}^2-\tila_{\alpha}^2\right)^{\upind{\gamma} N} \,, \\[0.0mm]
	U_{\alpha,0} & \approx \tilU_{\alpha} \prod_{\gamma} \left(\tila_{\gamma}^2-\tilx_{\alpha}^2\right)^{\upga N} \,,\\[0.0mm]
	U_{\alpha,\rho} & \approx \tilJ(\tila_{\alpha}^2) \upal U_{\rho} \left(2\tila_{\alpha}\varepsilon\right)^{\upal N-1} \prod_{\overset{\scriptstyle{\gamma}}{\scriptstyle{\gamma\neq\alpha}}} \left(\tila_{\gamma}^2-\tila_{\alpha}^2\right)^{\upga N} \,,\\[0.0mm]
	\calU_{\alpha,0} & \approx \tilcalU_{\alpha} \upal\calJ(0) \left(2\tila_{\alpha}\varepsilon\right)^{\upal N} \prod_{\overset{\scriptstyle{\gamma}}{\scriptstyle{\gamma\neq\alpha}}} \left(\tila_{\gamma}^2-\tila_{\alpha}^2\right)^{\upga N} \,,\\[0.0mm]
	\calU_{\alpha,\rho} & \approx -\tilcalU_{\alpha} \upal a_{\rho}\upal\calU_{\rho} \left(2\tila_{\alpha}\varepsilon\right)^{\upal N} \prod_{\overset{\scriptstyle{\gamma}}{\scriptstyle{\gamma\neq\alpha}}} \left(\tila_{\gamma}^2-\tila_{\alpha}^2\right)^{\upga N} \,.
\end{align}
The functions after the limit, such as $\tilJ$, $\tilA$, $\tilU$ and $\upal J$, $\upal A$, $\upal U$ are defined similarly to the functions before the limit in Appendix \ref{app:metric-funcs-def}, only the sets of coordinates and parameters they include are restricted to $\tilx_{\alpha}$, $\tila_{\alpha}$ and $\upal x_{\rho}$, $\upal a_{\rho}$, respectively. Moreover, the latter are defined using first powers of variables instead of their squares. For example, the definitions \eqref{eq:app-Jmu} and \eqref{eq:app-Amu} are modified as
\begin{equation}
\begin{aligned}
	\tilJ_{\alpha}(\tila^2) & = \prod_{\overset{\scriptstyle{\beta}}{\scriptstyle{\beta\neq\alpha}}} \left(\tilx_{\beta}^2-\tila^2\right) = \sum_r \tilA_{\alpha}^{(r)}(-\tila^2)^{\tilN-1-r} \,,\quad &
	\upal J_{\rho}(\upal a) & = \upal\prod_{\overset{\scriptstyle{\sigma}}{\scriptstyle{\sigma\neq\rho}}} \left(\upal x_{\sigma}-\upal a\right) = \sum_p \upal A_{\rho}^{(p)}(-\upal a)^{\upal N-1-p} \,, \\[0.0mm]
	\tilcalJ_{\alpha}(\tilx^2) & = \prod_{\overset{\scriptstyle{\beta}}{\scriptstyle{\beta\neq\alpha}}} \left(\tila_{\beta}^2-\tilx^2\right) = \sum_r \tilcalA_{\alpha}^{(r)}(-\tilx^2)^{\tilN-1-r} \,,\quad &
	\upal\calJ_{\rho}(\upal x) & = \upal\prod_{\overset{\scriptstyle{\sigma}}{\scriptstyle{\sigma\neq\rho}}} \left(\upal a_{\sigma}-\upal x\right) = \sum_p \upal\calA_{\rho}^{(p)}(-\upal x)^{\upal N-1-p} \,,
\end{aligned}
\end{equation}\\[-3ex]
where
\begin{equation}
\label{eq:app-A-alpharho}
\begin{aligned}
	\tilA_{\alpha}^{(r)} & = \mkern-18mu \sum_{\overset{\overset{\scriptstyle{\beta_1,\dots,\beta_r}}{\scriptstyle{\beta_1<\dots<\beta_r}}}{\scriptstyle{\beta_i\neq\alpha}}} \mkern-18mu \tilx_{\beta_1}^2\dots \tilx_{\beta_r}^2 \,,&\quad
	\upal A_{\rho}^{(p)} &= \mkern-18mu \sum_{\overset{\overset{\scriptstyle{\sigma_1,\dots,\sigma_p}}{\scriptstyle{\sigma_1<\dots<\sigma_p}}}{\scriptstyle{\sigma_i\neq\rho}}} \mkern-18mu \upal x_{\sigma_1}\dots \upal x_{\sigma_p} \,, \\[0.1cm]
	\tilcalA_{\alpha}^{(r)} & = \mkern-18mu \sum_{\overset{\overset{\scriptstyle{\beta_1,\dots,\beta_r}}{\scriptstyle{\beta_1<\dots<\beta_r}}}{\scriptstyle{\beta_i\neq\alpha}}} \mkern-18mu \tila_{\beta_1}^2\dots \tila_{\beta_r}^2 \,,&\quad
	\upal\calA_{\rho}^{(p)} &= \mkern-18mu \sum_{\overset{\overset{\scriptstyle{\sigma_1,\dots,\sigma_p}}{\scriptstyle{\sigma_1<\dots<\sigma_p}}}{\scriptstyle{\sigma_i\neq\rho}}} \mkern-18mu \upal a_{\sigma_1}\dots \upal a_{\sigma_p} \,,
\end{aligned}
\end{equation}\\[-3ex]
\endgroup
\end{widetext}
and the Latin indices go over the ranges
\begin{equation}
\begin{split}
	r & = 0, \ldots, \tilN-1 \,, \\
	p & = 0, \ldots, \upal N-1 \,.
\end{split}
\end{equation}
The other definitions are modified accordingly. These functions also satisfy analogous identities and orthogonality relations to \eqref{eq:id-A-sum-k}---\eqref{eq:og-JJcalJU}. Furthermore, we define $\upal J(\upal a)=1$ and $\upal\calJ(\upal x)=1$ if $\upal N=0$. In particular, in the Lorentzian case we have
\begin{equation}
	\upind{\tilN}J(\upal a) = 1 \,,\qquad \upind{\tilN}\calJ(\upal x) = 1 \,.
\end{equation}



%

\end{document}